\documentclass[abbrv,aps,prb,twocolumn,showpacs,preprintnumbers,amsmath,amssymb,
groupedaddress,nofootinbib]{revtex4}

\usepackage[ansinew]{inputenc}
\usepackage{graphicx}
\usepackage{dcolumn}
\usepackage{bm}

\def\Re{\hbox{Re}}
\def\Im{\hbox{Im}}
\def\Tr{\text{Tr}}

\newcommand{\diff}{\ensuremath{\text{d}}}

\newcommand{\up}{\ensuremath{\uparrow}}
\newcommand{\dn}{\ensuremath{\downarrow}}

\begin{document}
\title{Multi-orbital Kondo physics of Co in Cu hosts}
\author{Brigitte Surer}
\author{Matthias Troyer}
\author{Philipp Werner}
\affiliation{Theoretische Physik, ETH Zurich, 8093 Zurich, Switzerland}
\author{Andreas M. L\"auchli}
\affiliation{Max-Planck-Institut f\"ur Physik komplexer Systeme, N\"othnitzer Strasse 38, D-01187 Dresden, Germany}
\affiliation{Institut f\"ur Theoretische Physik, Universit\"at Innsbruck, Technikerstrasse 25/2, A-6020 Innsbruck, Austria}
\author{Tim O. Wehling}
\author{Aljoscha Wilhelm} 
\author{Alexander I. Lichtenstein}
\affiliation{1.\ Institut f\"ur Theoretische Physik, Universit\"at Hamburg,
D-20355 Hamburg, Germany}

\date{\today}

\begin{abstract}
We investigate the electronic structure of cobalt atoms on a copper surface and in a copper host by combining density functional calculations with a numerically exact continuous-time quantum Monte Carlo treatment of the five-orbital impurity problem. In both cases we find low energy resonances in the density of states of all five Co $d$-orbitals. The corresponding self-energies indicate the formation of a  Fermi liquid state at low temperatures. Our calculations yield the characteristic energy scale -- the Kondo temperature -- for both systems in good agreement with experiments. We quantify the charge fluctuations in both geometries and suggest that Co in Cu must be described by an Anderson impurity model rather than by a model assuming frozen impurity valency at low energies. We show that fluctuations of the orbital degrees of freedom are crucial for explaining the Kondo temperatures obtained in our calculations and and measured in experiments.
\end{abstract}
\maketitle

\section{Introduction}
The Kondo effect, arising when localized spins interact with a metallic environment, is a classic many body problem.\cite{HewsonBook} At present, idealized models dealing with, for instance,  a single spin degree of freedom screened by a sea of conduction electrons are well understood. 
Spin $S=1/2$ Kondo models or single orbital Anderson impurity models have been widely considered to describe magnetic impurities with open $d$-shells in metallic environments and proved helpful in qualitative discussions.\cite{Zawadowski_PRL00,Kern02,Wenderoth04,BJones_PRL06,Neel_08,Ternes_Heinrich_Schneider_Review09,Wahl_NJP09,Wenderoth11}
As realized, however, by Nozi\`{e}res and Blandin in 1980,\cite{NozieresBlandin_1980} such idealized models may ignore important aspects of the nature of transition metal impurities as they disregard orbital degrees of freedom. This makes comparisons between theory and experiment often very difficult. More realistic models accounting for the orbital structure, Hund's rule coupling, non-spherical crystal fields and an energy- and orbital-dependent hybridization of the impurity electrons with the surrounding metal are theoretically very demanding due to the multiple degrees of freedom and multiple energy scales involved. 

In recent years different attempts have been made to address this problem. For the classic example of Fe in Au, which has been experimentally studied since the 1930s, a model describing the low energy physics has been derived\cite{Costi_09} by comparing numerical renormalization group (NRG) calculations to electron transport experiments. The Kondo temperature $T_K$, below which the impurity spin becomes screened and a Fermi liquid develops, served as a fitting parameter in this study. A scaling analysis of multiple Hund's coupled spins in a metallic environment showed that Hund's rule coupling can strongly quench the formation of Kondo singlet states.\cite{Coleman_PRL09} For highly symmetric systems like Co adatoms on graphene\cite{Co_graphene_Kondo_PRB10} or Co-benzene sandwich molecules in contact to metallic leads,\cite{Karolak_11} the orbital degree of freedom has been suggested to control Kondo physics down to the lowest energy scale. However, a general strategy to assess which degrees of freedom are involved in the formation of low energy Fermi liquids around magnetic impurities in metals is still lacking.

Co atoms coupled to Cu hosts present another experimentally extensively studied system which has been interpreted in terms of Kondo physics.\cite{Manoharan_00,Kern02,Wenderoth04,Neel_08,Ternes_Heinrich_Schneider_Review09,Wahl_NJP09,Wenderoth11}
Theoretical descriptions of this system have  often been based on single orbital Anderson impurity models \cite{Zawadowski_PRL00,Kern02,Wenderoth04,BJones_PRL06} or Kondo models\cite{Neel_08} and the role of orbital fluctuations in these systems has remained rather unclear. Recently developed continuous time Quantum Monte Carlo (CTQMC) \cite{Gull:2011} approaches allow to describe the full orbital structure of magnetic impurities in metallic hosts, while accounting for all electron correlations in a numerically exact way. So far, however, such CTQMC studies have been limited to rather high temperatures,\cite{Gorelov_PRB09} well above typical Kondo scales on the order of 10K to 500K. 
The realistic description of transition metal Kondo systems thus remains a long standing open problem in solid state physics.

\begin{figure}
\includegraphics[width=.8\linewidth]{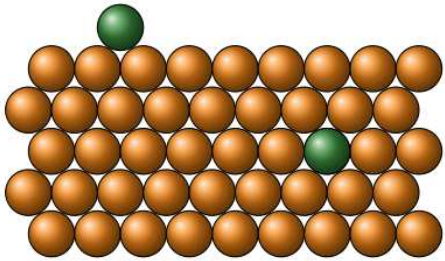} 
\caption{\label{sketch}(Color online)
Sketch of Co impurities in a Cu host. We consider the two cases: (i) the impurity buried in the bulk (Co in Cu), and (ii) on top of the Cu layer (Co on Cu).
}
\label{fig:imp}
\end{figure}

Here, we employ the recently developed Krylov CTQMC method \cite{Laeuchli:2009p653} in combination with density functional based first-principles calculations to achieve an ab-initio description of two archetypical Kondo systems: Co adatoms on a Cu (111) surface, as well as Co impurities in bulk Cu (see Fig. \ref{fig:imp}). We consider the energy dependent hybridization of the impurities with the surrounding host material as well as the full local Coulomb interaction and find low energy resonances developing in the spectral function as the temperature is lowered. Such resonances are found in all impurity $3d$-orbitals and our calculations indicate that spin- and orbital-fluctuations are crucial for the formation of low energy Fermi liquids involving all impurity $3d$-orbitals.
We also demonstrate the intermediate-valence character of the Co impurity in bulk, which implies that the physics cannot be correctly described by a low-energy Kondo model which neglects charge fluctuations.

The paper is organized as follows. Section \ref{sec:model} defines the model we use to study the Kondo physics in Co coupled to Cu hosts. Section \ref{sec:computationalmethod} specifies the Density Functional Theory calculations of the hybridization function (\ref{sec:densityfunctionalcalculations}) and describes the Krylov CTQMC method used to solve the five-orbital impurity problem (\ref{sec:impuritysolver}). In Section \ref{sec:results} we present the calculated quasi-particle spectra (\ref{sec:quasiparticlespectra}), exemplify the low energy Fermi-liquid behavior based on the self-energies (\ref{sec:lowenergyfermiliquid}) and estimate the Kondo temperature from our Quantum Monte Carlo data (\ref{sec:estimationoftk}). In Section \ref{scalepic} we discuss the prevalence of charge fluctuations and the implications of their presence (\ref{sec:chargeflucuations}). We elaborate on the role of spin and orbital fluctuations and their influence on the low energy behavior (\ref{sec:spinandorbitalflucations}). Section \ref{sec:conclusions} is a summary and conclusion emphasizing the implications of our findings and outlining prospects for future investigations.

\section{Model}
\label{sec:model}
A realistic description of the Co atoms on Cu (111) and in bulk Cu including all five Co $3d$ orbitals can be formulated in terms of a multi-orbital Anderson impurity model: 
\begin{equation}
\label{eqn:H_AIM}
H_{\rm AIM}=\sum_{k}\epsilon_k c^\dagger_{k}c_{k}+\sum_{k,m}  \left(V_{km} c^\dagger_{k}d_{\alpha} + \rm{H.c.}\right)+H_{\rm loc}
\end{equation}
with
\begin{equation}
 H_{\rm loc}=\sum_\sigma \epsilon_{\alpha}d^\dagger_{\alpha}d_{\alpha}+ \frac{1}{2}\sum_{\alpha_1,...,\alpha_4} U_{\alpha_1,...,\alpha_4} d_{\alpha_1}^\dagger d_{\alpha_2}^\dagger d_{\alpha_3} d_{\alpha_4}.
\end{equation}

It describes an impurity characterized by quantum numbers $\alpha$ (orbital and spin), with corresponding annihilation operators $d_{\alpha}$, on-site energies $\epsilon_{\alpha}$, and local Coulomb interactions $U_{\alpha_1,...,\alpha_4}$. This impurity is embedded in a sea of conduction electrons described by annihilation operators $c_{k}$ and dispersions $\epsilon_k$, where $k$ includes crystal momentum, band index and spin. The coupling between the impurity and the conduction electrons is provided by the hybridization $V_{k\alpha}$.

In the AIM only the impurity site is subject to a quartic interaction term, whereas the bath of conduction electrons is assumed to be non-interacting.  The bath degrees of freedom can thus be integrated out and the local electronic properties of the impurity can be described by the effective action
\begin{widetext}
\begin{equation}
\label{eqn:Seff}
 S_{\rm eff}=-\sum_{\alpha_1,\alpha_2}\int_0^\beta\diff\tau\int_0^\beta\diff\tau'd^*_{\alpha_1}(\tau)\mathcal{G}_{0,\alpha_1\alpha_2}^{-1}(\tau,\tau')d^*_{\alpha_2}(\tau')+\int_0^\beta \diff\tau H_{\rm loc}(d^*_{\alpha_1}(\tau),d_{\alpha_2}(\tau))
\end{equation}
\end{widetext}
with
\begin{equation}
\label{eqn:bath_G}
 \mathcal{G}_{0,\alpha_1\alpha_2}^{-1}(i\omega_n)=(i\omega+\mu)-\Delta_{\alpha_1\alpha_2}(i\omega_n)
\end{equation}
and the hybridization function
\begin{equation}
\label{eqn:def_hybridization_function}
\Delta_{\alpha_1\alpha_2}(\omega)=\sum_{k}\frac{V_{k\alpha_1}^*V_{k\alpha_2}}{i\omega-\epsilon_k}.
\end{equation}
To specify the parameters of impurity models describing Co on Cu (111) as well as Co in bulk Cu we have performed first principles density functional theory calculations.

\section{Computational methods}
\label{sec:computationalmethod}

\subsection{Density functional calculations}
\label{sec:densityfunctionalcalculations}

Density functional theory (DFT) calculations were performed to obtain relaxed geometries and the hybridization functions for single Co atoms in Cu and on a Cu (111) surface. The DFT calculations have been carried out using a generalized gradient approximation (GGA) \cite{Perdew:PW91} as implemented in the Vienna Ab-Initio Simulation Package (VASP) \cite{gkr_94} with  projector augmented waves basis sets (PAW).\cite{gkr_99,pbl_94} For the simulation of a cobalt impurity in bulk Cu we employed a CoCu$_{63}$ supercell structure. Co on Cu (111) was modeled using a $3\times 4$ supercell of a Cu (111) surface with a thickness of 5 atomic layers and a Co adatom on the surface, see Fig. \ref{sketch}. All structures were relaxed until the forces acting on each atom were below $0.02\,$eV\AA$^{-1}$. For Co in bulk Co the entire supercell and for Co on Cu (111) the adatom and the three topmost Cu layers were relaxed.
The PAW basis sets provide intrinsically projections onto localized atomic orbitals, which we used to extract the hybridization functions (Eq.~(\ref{eqn:def_hybridization_function})) from our DFT calculations (for details see Refs.~\onlinecite{PAW_DMFT,DFT++}). 

\subsection{Impurity solver}
\label{sec:impuritysolver}

The impurity model (\ref{eqn:Seff}) can be solved without approximations using  continuous-time quantum Monte Carlo (CTQMC) algorithms. Both the interaction expansion \cite{Rubtsov:2005p1699} and hybridization expansion \cite{Werner:2006p11,Werner:2006p2} algorithms can treat multi-orbital systems with general four-fermion interaction terms.  The hybridization expansion is advantageous in the case of a strongly interacting five orbital model, because the expansion in the hybridization leads to much lower perturbation orders than an expansion in the various interaction terms, and thus to much reduced computational demands. For the interaction parameters of the Co impurities studied in this paper the order in the hybridization expansion was found to be a factor $20$ lower compared to the interaction  expansion.\cite{Gorelov_PRB09} Furthermore, if the hybridization function is diagonal in the orbital indices, as it is to a good approximation the case in the Co/Cu systems studied here, no sign problem appears. This is in contrast to the hybridization expansion, where correlated hopping terms were found to lead to severe sign cancellations.\cite{Gorelov_PRB09} 

Despite these advantages, a strong-coupling CTQMC simulation of a general five orbital model is still computationally expensive, because the imaginary time evolution of the local impurity Hamiltonian $H_\text{loc}$ has to be computed exactly. The weight of a Monte Carlo configuration corresponding to $2n$ hybridization events (impurity creation and annihilation operators $d^\dagger_\alpha$ and $d_\alpha$) at times $\tau'_1<\ldots<\tau'_n$ and $\tau_1<\ldots<\tau_n$ contains a factor \cite{Werner:2006p2}
\begin{equation}
\Tr_\text{loc} \Big[e^{-\beta H_\text{loc}} T 
d_{\alpha_{n}}(\tau_n)d^\dagger_{\alpha_n'}(\tau_{n}')
\ldots
d_{\alpha_1}(\tau_1)d^\dagger_{\alpha_1'}(\tau_1') 
\Big],
\label{trace}
\end{equation}
where the trace and the operator products operate on a  Hilbert space of $H_\text{loc}$ of dimension 1024. One strategy to evaluate this trace is to use the eigenbasis of $H_\text{loc}$, in which the time-evolution operators are diagonal, and to order the eigenstates according to conserved quantum numbers. The operators $d_\alpha^{(\dagger)}$ then acquire a block structure \cite{Haule:2007p693} which reduces the effort for matrix-matrix multiplications. 

An alternative approach, which was found to be more efficient in the case of five orbital systems and low temperature, is the Krylov implementation.\cite{Laeuchli:2009p653} In this algorithm one evaluates the trace factor in the occupation number basis. In this basis, the operators $d_\alpha^{(\dagger)}$ are sparse and can easily be applied to any state, while the time-evolution operators become non-trivial dense matrices. However, we never evaluate the exponential of $H_\text{loc}$, but only $\exp(-H_\text{loc}\tau)|v\rangle$, for a given state $|v\rangle$. This can be done with only a small number of sparse matrix-vector multiplications, by using efficient Krylov space techniques. Details of the algorithm can be found in Ref.~\onlinecite{Laeuchli:2009p653}. The computational advantage of this Krylov-space approach becomes particularly evident at low temperatures, where it is sufficient to consider the contributions of the low energy states in $\Tr_\text{loc}[\ldots]$, since the probability of the system to be in one of these states is very large (during the time evolution, all excited states are accessible). If the outer trace is truncated in this way, the observables must be measured in the middle of the imaginary-time interval, i.e. at $\tau=\beta/2$. 

\section{Results}
\label{sec:results}

\begin{figure}%
\begin{tabular}{ll}
a)&b)\\
\includegraphics[width=.49\columnwidth]{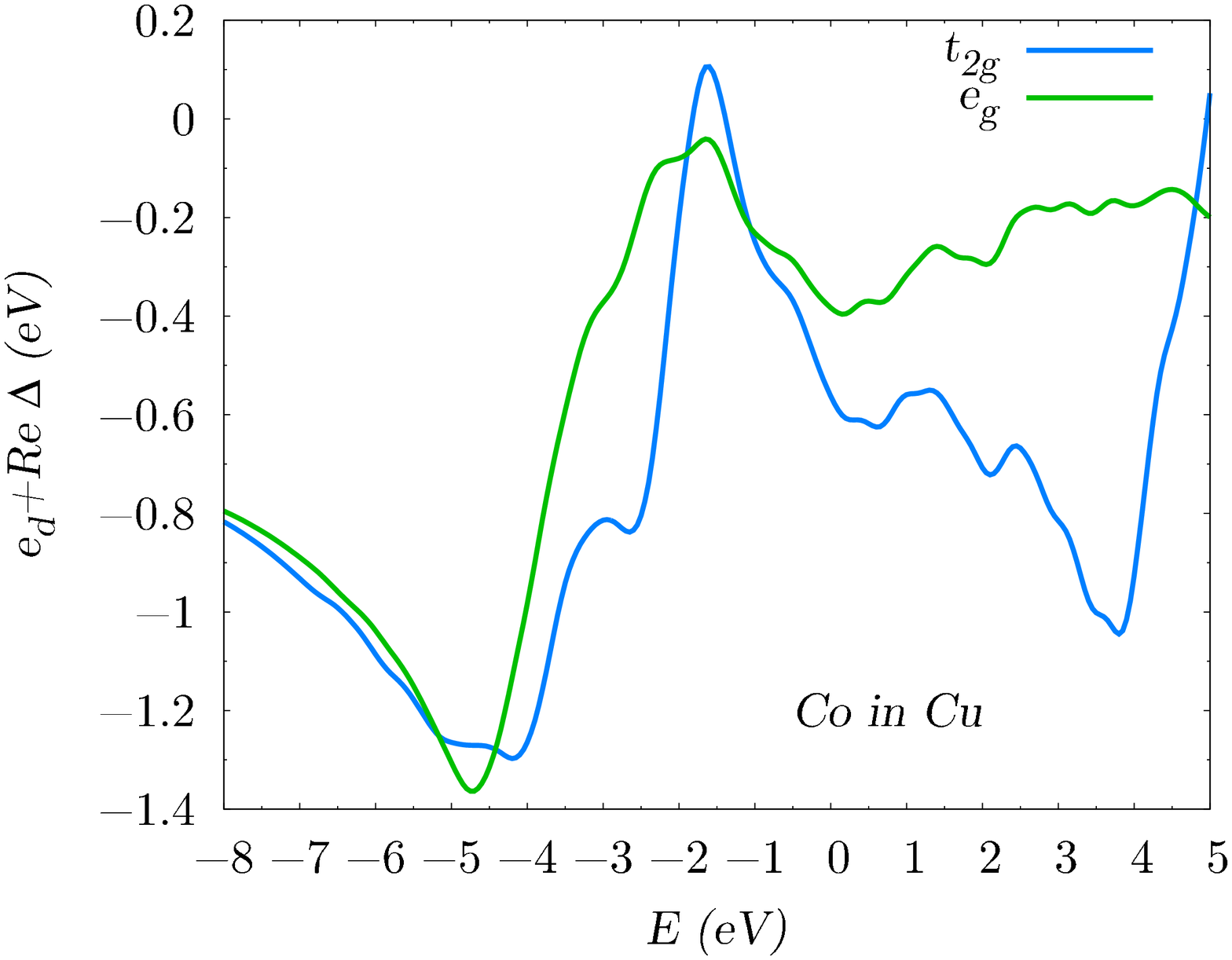}&\includegraphics[width=.49\columnwidth]{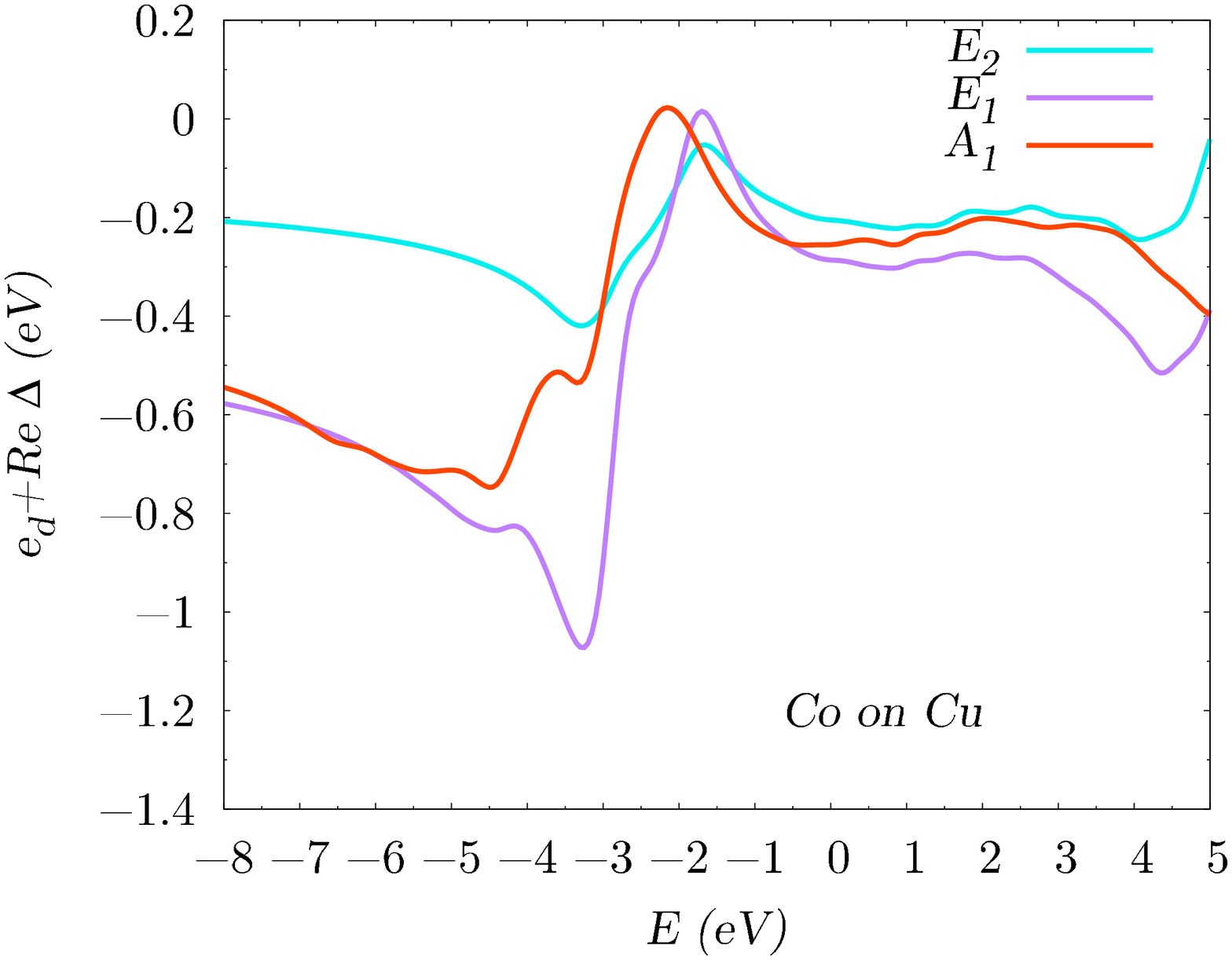}\\
\includegraphics[width=.49\columnwidth]{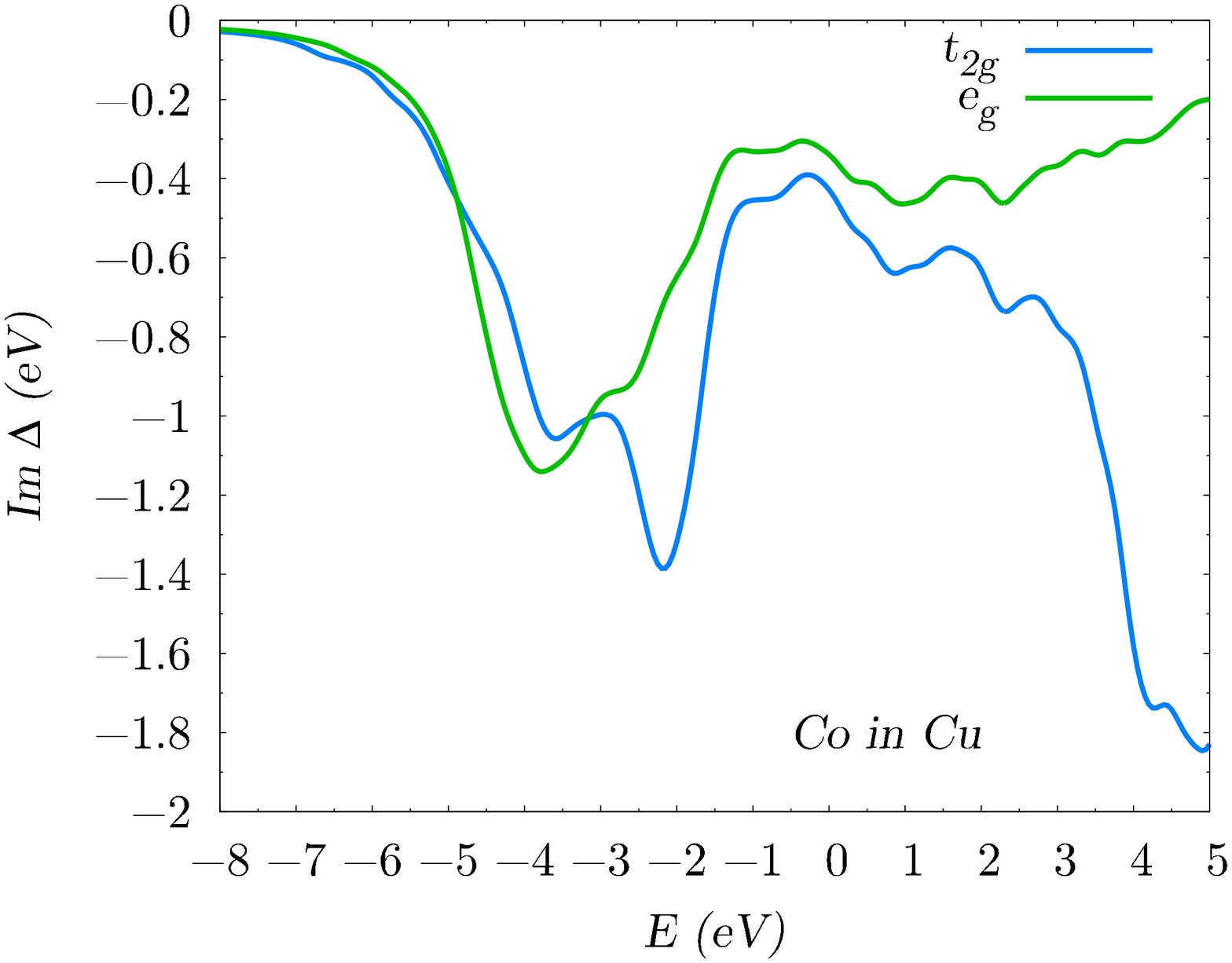}&\includegraphics[width=.49\columnwidth]{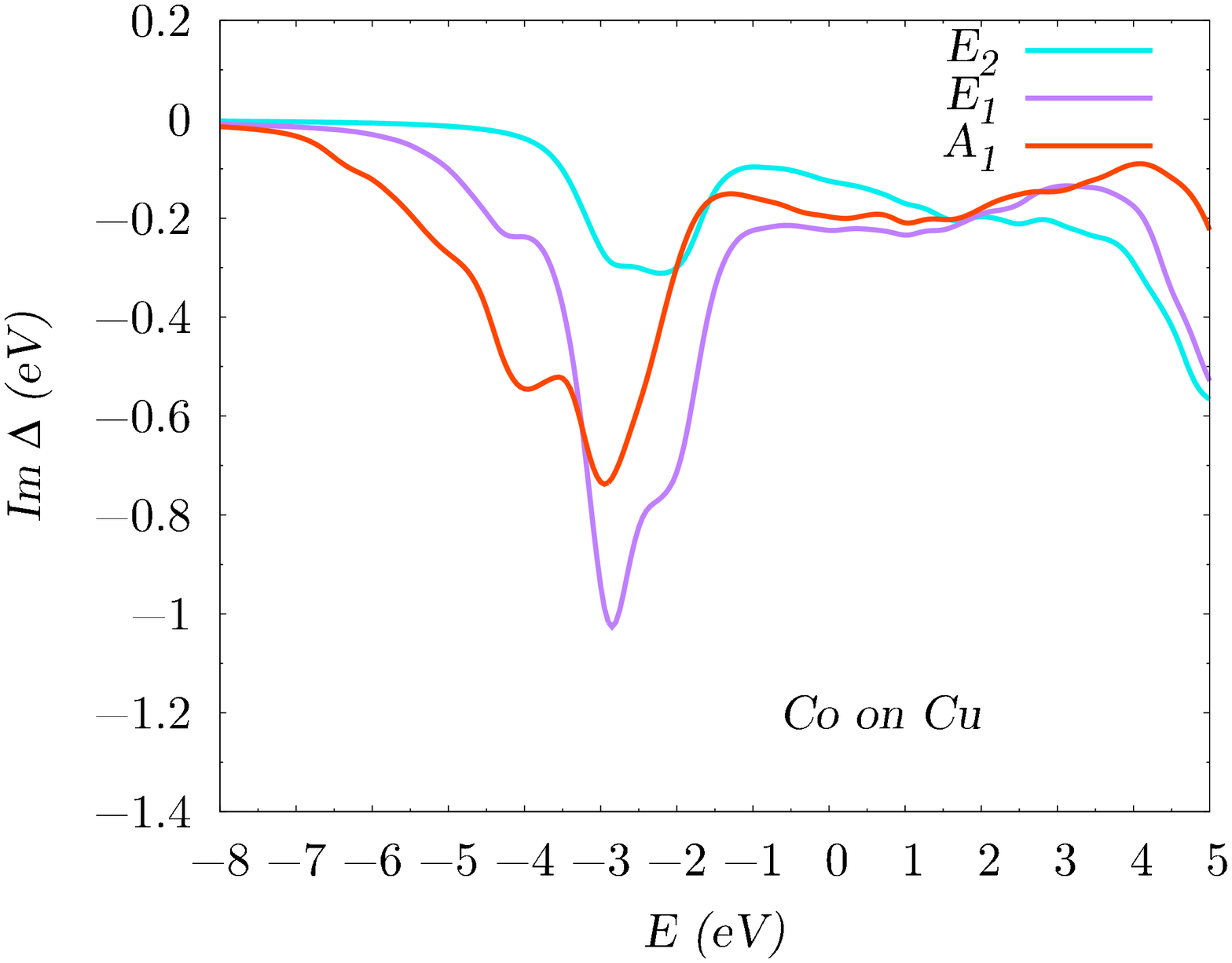}
\end{tabular}
\caption{(Color online) Hybridization functions and crystal fields for Co in bulk Cu (a) and as ad-atom on Cu (111) (b). In the upper panel the dynamical crystal field $\epsilon_d+\Re\,\Delta$ is shown. Lower panel: $\Im\,\Delta$.
}%
\label{fig:DFT_hybridization}%
\end{figure}

The DFT calculations yield the orbital dependent hybridization functions shown in Fig. \ref{fig:DFT_hybridization}. In bulk Cu, the environment of the Co impurities is cubically symmetric and the hybridization function decomposes into  threefold degenerate $t_{2g}$ and  twofold degenerate $e_g$ blocks. In the bulk symmetry forbids off-diagonal elements in the hybridization function. On the surface, the symmetry is reduced to $C_{3v}$. For Co on Cu the hybridization function decomposes into two twofold degenerate blocks transforming according to the E-irreducible representation of $C_{3v}$ ($E_1$ ($d_{xz}$, $d_{yz}$) and $E_2$ ($d_{x^2-y^2}$, $d_{xy}$)) and the $d_{z^2}$-orbital transforming according to the $A_1$ representation. For Co on Cu the hybridization functions contain small off-diagonal matrix elements. These off-diagonal elements, which are smaller than the diagonal ones,  will be neglected in our simulations. 
As a general trend, one can already see that the hybridization of the Co $d$-electrons is about twice larger in the bulk than on the surface. 

DFT calculations are also used to calculate the occupancy of the Co $3d$ impurity orbitals. To this end, we performed spin-polarized DFT calculations using GGA as well as GGA+U of Co in and on Cu with the full interaction vertex defined via the average screened Coulomb interaction $U=4\,$eV and the exchange parameter $J=0.9$\,eV. We obtained the occupancies of the Co $3d$ orbitals derived from the PAW projectors $n=n_\up+n_\dn$ and the impurity spin $S_z=\frac{1}{2}(n_\up-n_\dn)$ and present them in Tab. \ref{tab:DFT_occs}. 
In all cases the average Co $3d$ occupancy suggested by our DFT calculations is between $n=7$ and $n=8$. For Co on Cu, the impurity spin is $S_z\approx 1$ which is well in line with a $d^8$ configuration of the Co. In the bulk, the Co spin is $S_z\approx 1$ in GGA+U and $S_z\approx 1/2$ in GGA.

\begin{table}%
\begin{tabular}{|l|c|c|c|c||c|c|c|c|}
\hline
&\multicolumn{4}{c||}{GGA}&\multicolumn{4}{c|}{GGA+U}\\
\hline
&$n$&$S_z$&$\tilde n$&$\tilde S_z$&$n$&$S_z$&$\tilde n$&$\tilde S_z$\\
\hline
Co on Cu & 7.3 & 0.96 & 7.6 & 1.00 & 7.4 & 0.96 & 7.7 & 1.00 \\
Co in Cu & 7.3 & 0.51 & 7.6 & 0.53 & 7.3 & 0.90 & 7.5 & 0.93 \\
\hline
\end{tabular}
\caption{Occupancies $n$ and impurity spins $S$ as obtained from our GGA and GGA+U calculations. Values obtained directly from the PAW projectors ($n,S$) and normalized by the integrated total Co $d$-electron DOS, $\mathcal{N}=\int \nu(E)\diff E$, are shown ($\tilde n=n/\mathcal{N}$, $\tilde S_z=S_z/\mathcal{N}$).}
\label{tab:DFT_occs}
\end{table}

In the following we study Co in and on Cu in the five-orbital Anderson impurity model formulation (Eq.~(\ref{eqn:H_AIM})). In this framework, the chemical potential has to be chosen to fix the occupancy of the Co $d$-orbitals. Due to the well know double-counting problem in LDA+DMFT type approaches,\cite{PAW_DMFT} the precise chemical potential $\mu$ and the Co $d$-occupancy are not known. Therefore, we computed results in a range of chemical potential values which yield a total $d$ occupancy consistent with the estimates of the DFT calculations. For both systems the results of the DFT calculations predict a total density $n\lesssim 8$ and suggest a spin $S\approx 1$ or slightly below in the case of Co in Cu. 
For $\mu=26,27,28$\,eV (Co in Cu) and $\mu=27,28,29$\,eV (Co on Cu) we obtain total densities and spins close to these DFT estimates. The values  of both observables for the lowest simulation temperature $T=0.025$\,eV are presented in Table \ref{tab:sn}.

\begin{table}[b]
\caption{
Total density and spin.
\label{tab:sn}}
{\footnotesize
\begin{tabular*}{\columnwidth}{@{\extracolsep{\fill}} l@{\hspace{-2.25em}}c c c c}
\hline
\hline
System & $\mu$ (eV) & $\langle n\rangle$  &  $\langle S\rangle$  \\
\hline
Co in Cu &$26$ &  $7.51\pm0.07$ &$1.02\pm0.02$   \\
Co in Cu &$27$ &  $7.78\pm0.05$ &$0.92\pm0.02$   \\
Co in Cu &$28$ &  $8.06\pm0.03$ &$0.817\pm0.007$   \\
\hline
Co on Cu & $27$ &  $7.76\pm0.05$ &$1.07\pm0.01$   \\
Co on Cu & $28$ &  $7.93\pm0.05$ &$0.99\pm0.01$   \\
Co on Cu & $29$ &  $8.21\pm0.03$ &$ 0.860\pm0.007$   \\
\hline
\hline
\end{tabular*}
}
\end{table}

\subsection{Quasi-particle spectra}
\label{sec:quasiparticlespectra}
\begin{figure}%
\begin{minipage}{1.\linewidth}
\includegraphics[width=\columnwidth]{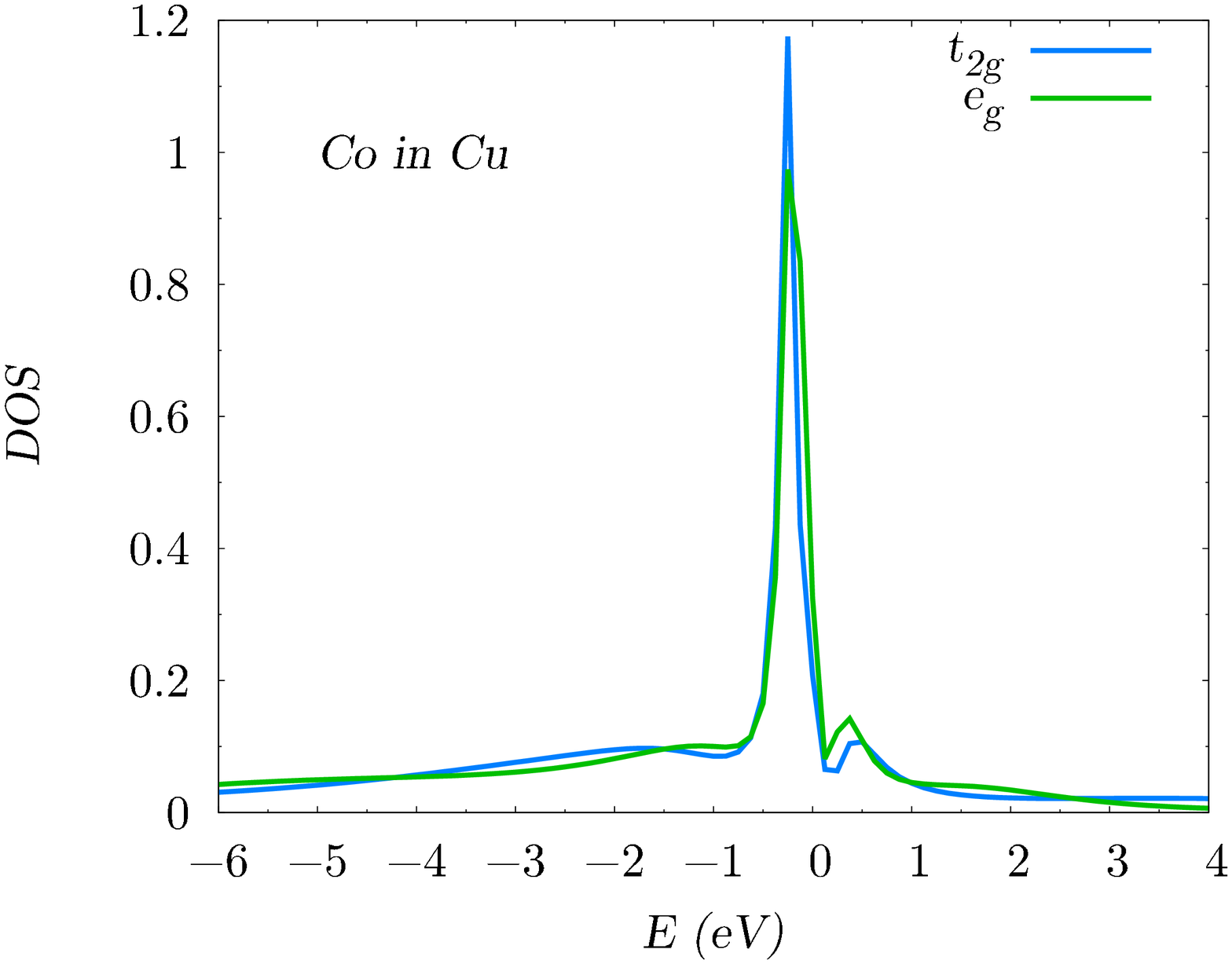}
\end{minipage}
\begin{minipage}{1.\linewidth}
\includegraphics[width=\columnwidth]{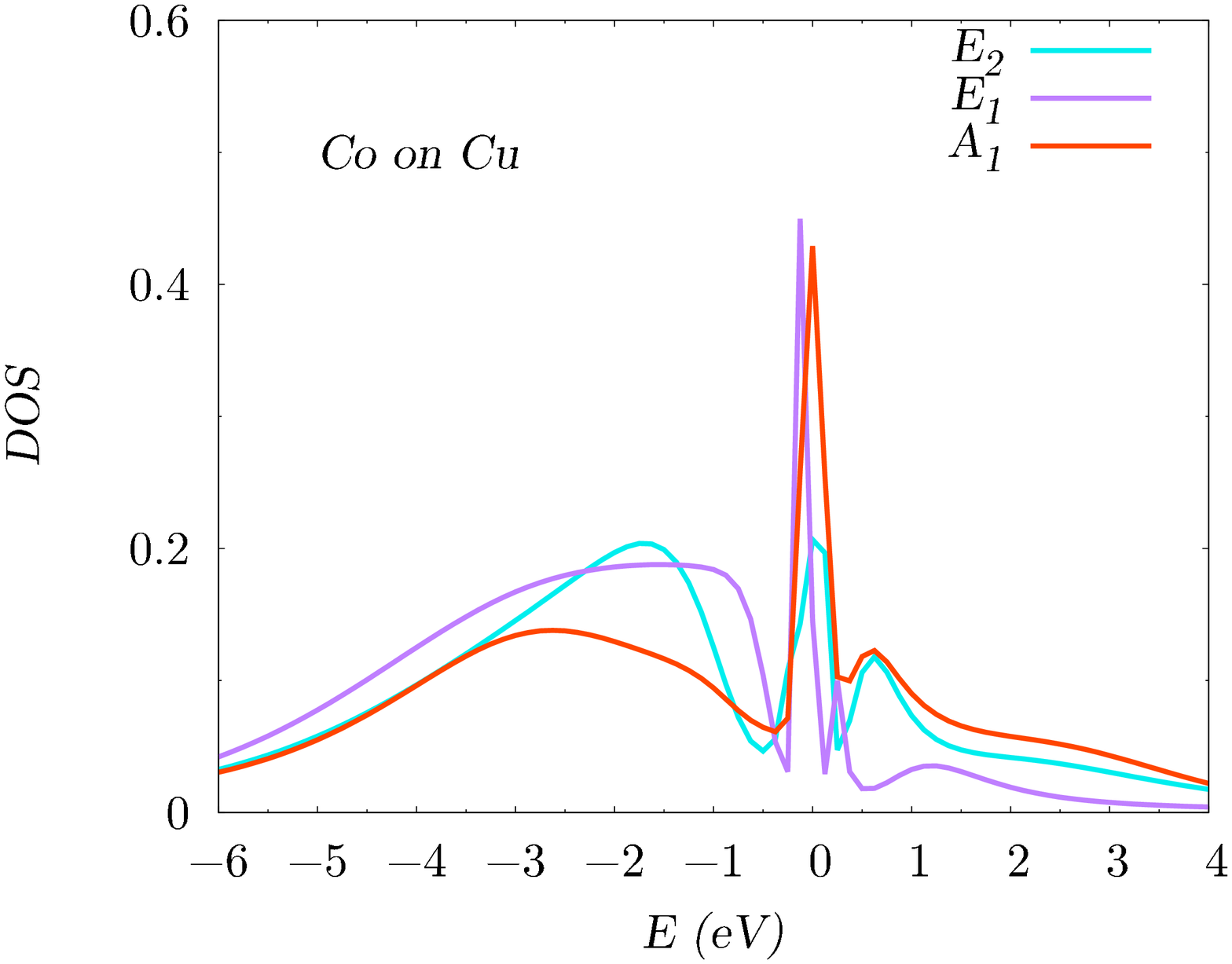}
\end{minipage}
\caption{(Color online) Orbitally resolved DOS of the Co impurities in bulk Cu (top) and on Co (111) (bottom) obtained from our QMC simulations at temperature $T=0.025$\,eV and chemical potential  $\mu=27$\,eV and $\mu=28$\,eV, respectively.}%
\label{fig:DOS_QMC_orbresolved}%
\end{figure}

\begin{figure}%
\begin{minipage}{1.\linewidth}
\includegraphics[width=\columnwidth]{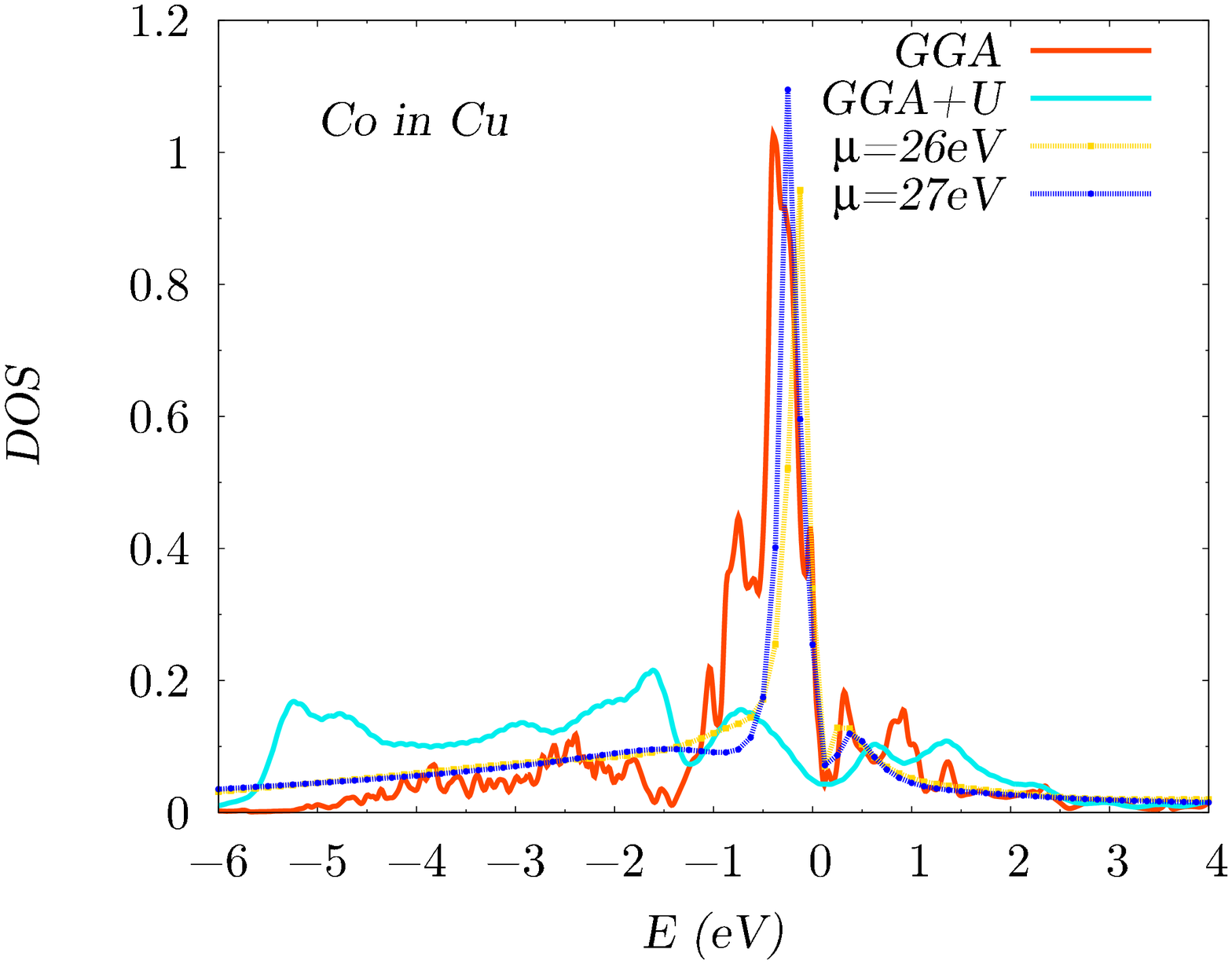}
\end{minipage}
\begin{minipage}{1.\linewidth}
\includegraphics[width=\columnwidth]{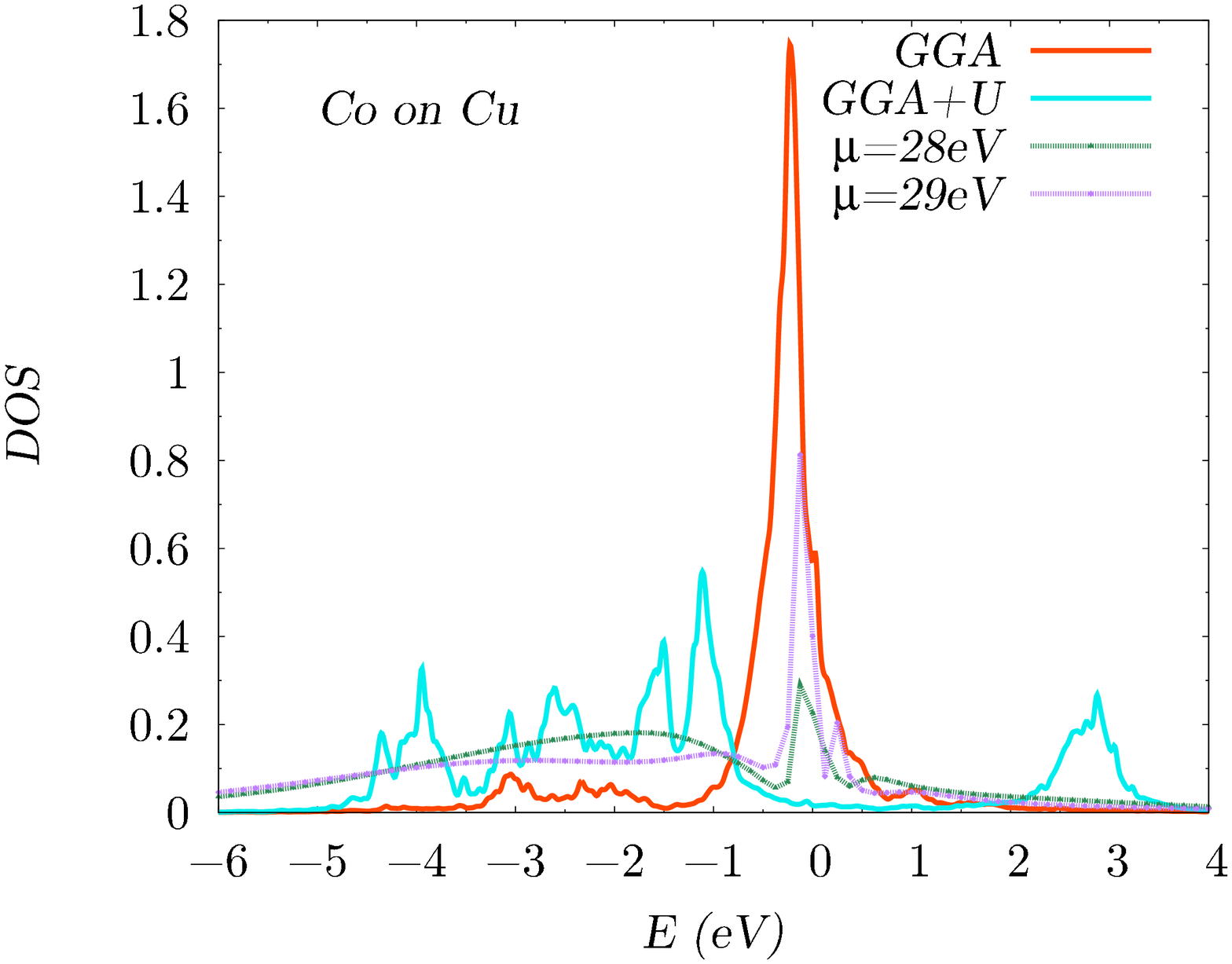}
\end{minipage}
\caption{(Color online) DOS of the Co impurities in bulk Cu (top) and on Cu (111) (bottom) obtained from DFT (GGA and GGA+U) as well as QMC simulations at temperature $T=0.025$\,eV. QMC results obtained at different chemical potentials $\mu$ are shown.}%
\label{fig:DOS_DFT_QMC}%
\end{figure}

We now analyze the excitation spectra of the Co impurities in order to understand the dominant physics at different energy scales. For a first, qualitative insight into the strength of many body renormalizations, we compare in Fig. \ref{fig:DOS_DFT_QMC} the Co $3d$-electron DOS obtained from our DFT calculations to the Co $3d$ spectral functions obtained from analytical continuation of our QMC results.

The non-spinpolarized GGA calculations used to determine the hybridization functions yield --- by definition --- the LDOS corresponding to the Anderson model without two-particle interactions ($U=J=0$\,eV). For both, Co in and on Cu, the GGA DOS exhibits a peak near the Fermi level ($E_F=0$). The QMC DOS qualitatively reproduces the GGA DOS for the case of Co in Cu. Here, the main difference between the two approaches is that QMC yields a peak near the Fermi level which is approximately twice narrower and shifted towards $E_F$. GGA+U accounts for the local Coulomb interactions at the Co atoms on a Hartree Fock level which leads to the destruction of the quasi-particle peak near $E_F$ with all the spectral weight shifted to broad Hubbard bands. The comparison to the QMC results shows that this destruction of the quasi-particle peak is unphysical.

For Co on Cu the hybridization is weaker and the DOS from the QMC simulations exhibits both quasi-particle peaks near $E_F$ as well as Hubbard type bands at higher energies. The reduction of spectral weight of the quasi-particle peak as compared to GGA is stronger here.

The orbitally resolved DOS of Co in and on Co is shown in Fig. \ref{fig:DOS_QMC_orbresolved}. For Co in Cu the DOS of the $e_g$ and the $t_{2g}$ orbitals is very similar particularly regarding the quasi-particle peak --- despite the (energy dependent) crystal field splitting on the order of some $0.1$\,eV.

The DOS of Co on Cu exhibits stronger differences between the $E_1$, $E_2$, and $A_1$ orbitals. The $E_2$ orbitals, which spread out perpendicular to the $z$-axis, show the weakest hybridization effects, but even here, a quasi-particle peak appears in all orbitals. The appearance of low energy quasi-particle peaks in all orbitals is different from the behavior expected for a spin-1 two-channel Kondo model, where a low energy quasi-particle resonance would be observed in two orbitals (four spin-orbitals) only.

The DOS as obtained from our QMC calculations suggests a low-temperature Fermi liquid state involving all orbitals for both, Co in and on Cu. We investigate the nature of this state in the following sections by analyzing the self-energies obtained from QMC and the statistics of relevant atomic states.

\subsection{Low energy Fermi liquid}
\label{sec:lowenergyfermiliquid}
If a Fermi liquid develops, the self-energy takes the form 
\begin{equation}
\Sigma(T,\omega)=\Sigma(T,0)+\Sigma'(T,0)\omega+O(\omega^2)
\label{eq:sigma_fl}
\end{equation} with $\Sigma(T,0)$ and the first energy derivative $\Sigma'(T,0)$ being real for $T\to 0$. In this regime, the spectral weight $Z$ associated with the quasi-particle peak is determined by
\begin{equation}
\label{eq:def_Z}
Z=\left(1-\text{Re}\Sigma'(0)- \text{Re}\Delta'(0)\right)^{-1}.
\end{equation}
Our QMC calculations yield the self-energy on the Matsubara axis. Analytic continuation $\omega\to i\omega_n$ shows that Fermi liquid behavior manifests itself on the Matsubara axis by 
\begin{equation}
\Im \Sigma(T,i\omega_n)\approx\Im\Sigma(T,0)-\Im\Sigma'(T,0)\omega_n
\label{eq:Im_Sigma_iomega}
\end{equation}
at low frequencies with $\Im\Sigma(T,0)\sim T^2$. 

\begin{figure*}%
\begin{minipage}{.32\linewidth}
\includegraphics[width=.99\columnwidth]{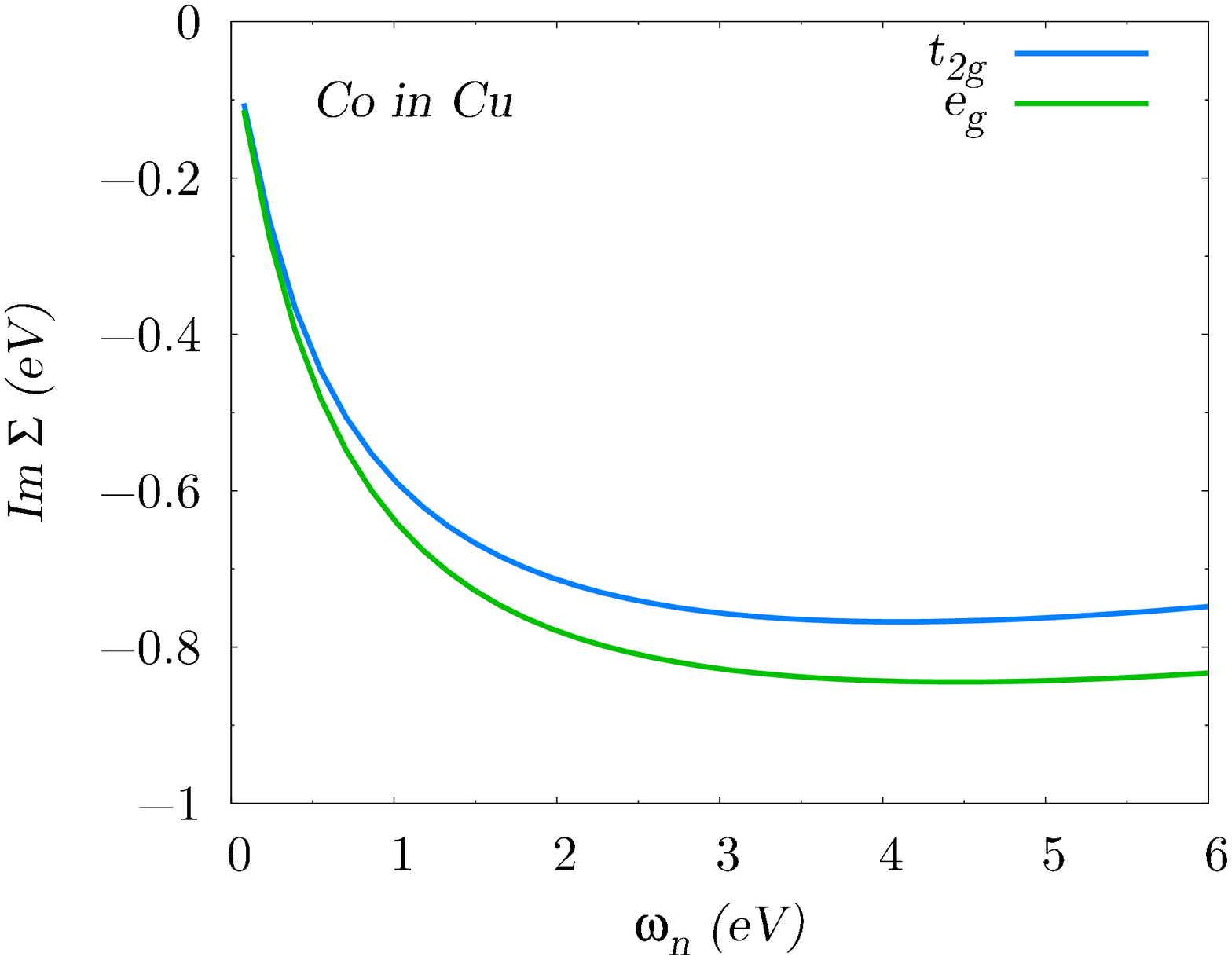}
\end{minipage}
\begin{minipage}{.32\linewidth}
\includegraphics[width=.99\columnwidth]{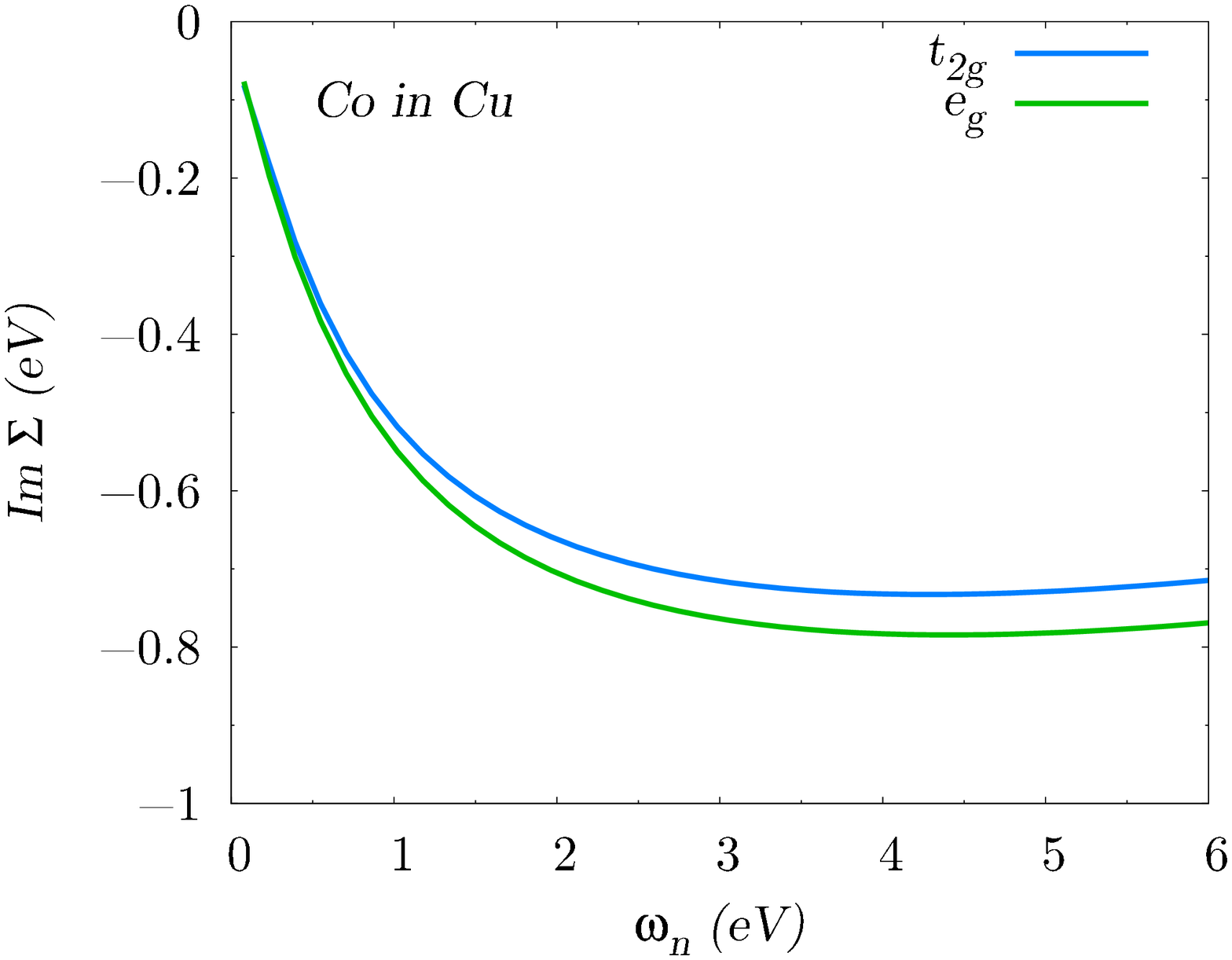}
\end{minipage}
\begin{minipage}{.32\linewidth}
\includegraphics[width=.99\columnwidth]{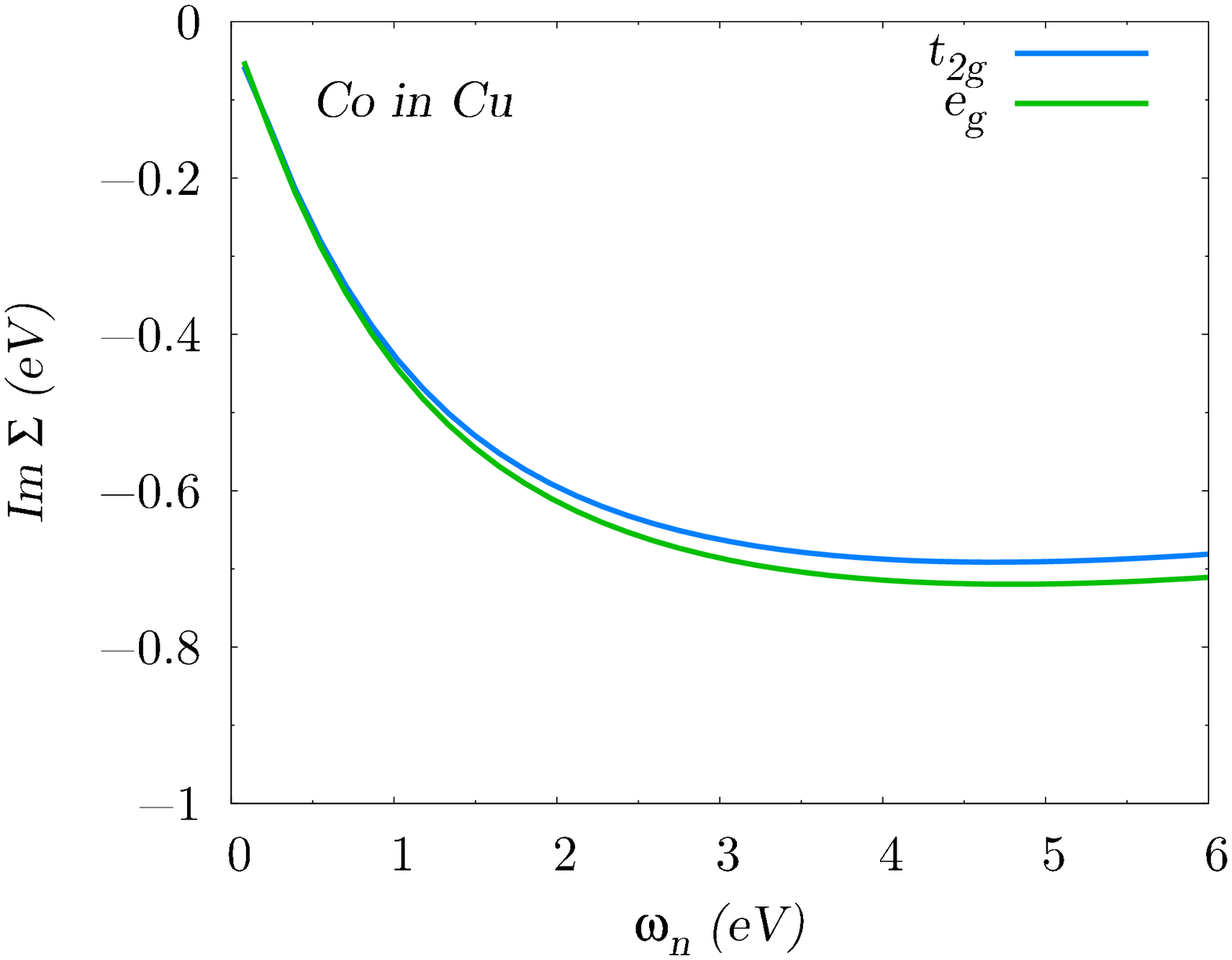}
\end{minipage}

\begin{minipage}{.32\linewidth}
\includegraphics[width=.99\columnwidth]{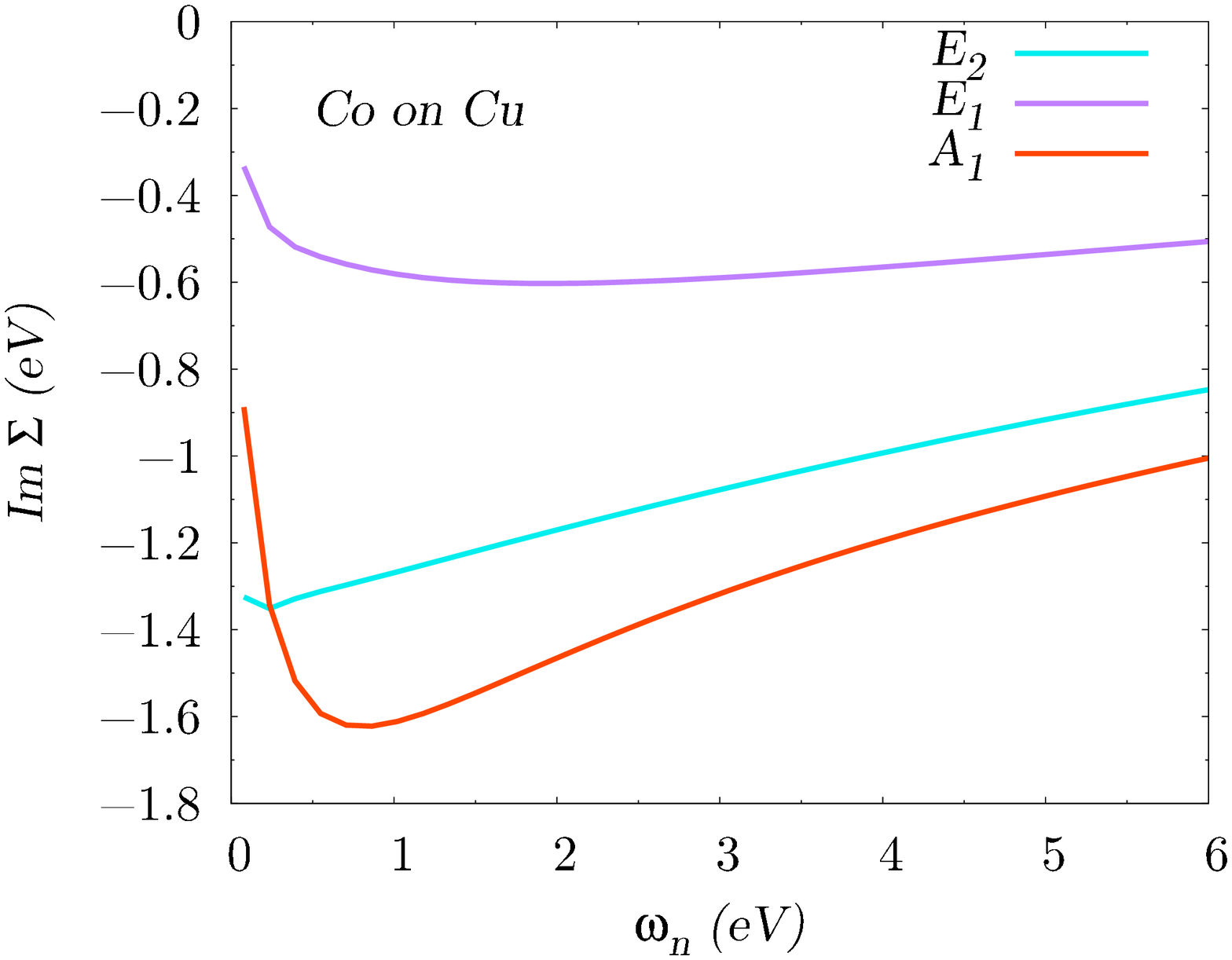}
\end{minipage}
\begin{minipage}{.32\linewidth}
\includegraphics[width=.99\columnwidth]{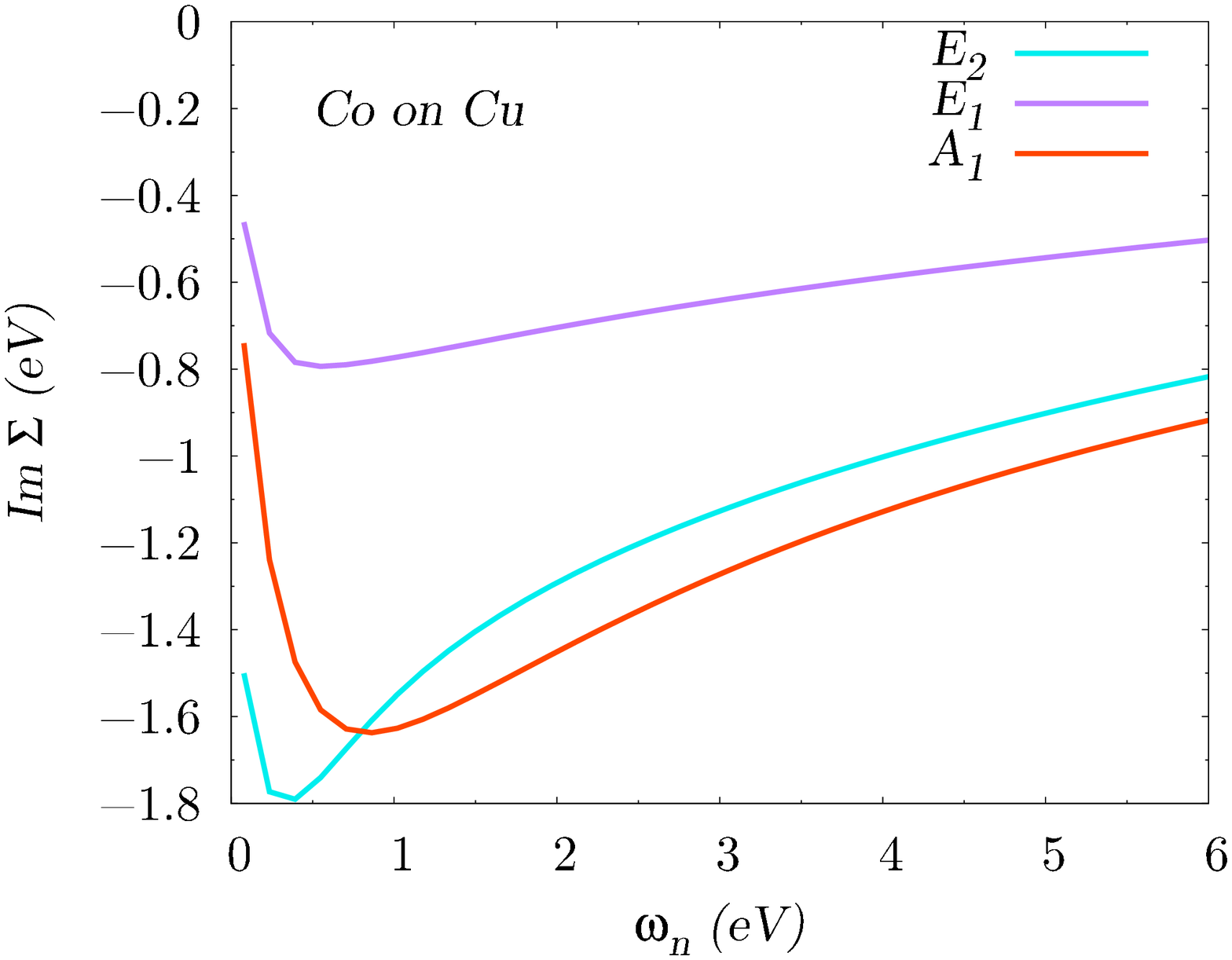}
\end{minipage}
\begin{minipage}{.32\linewidth}
\includegraphics[width=.99\columnwidth]{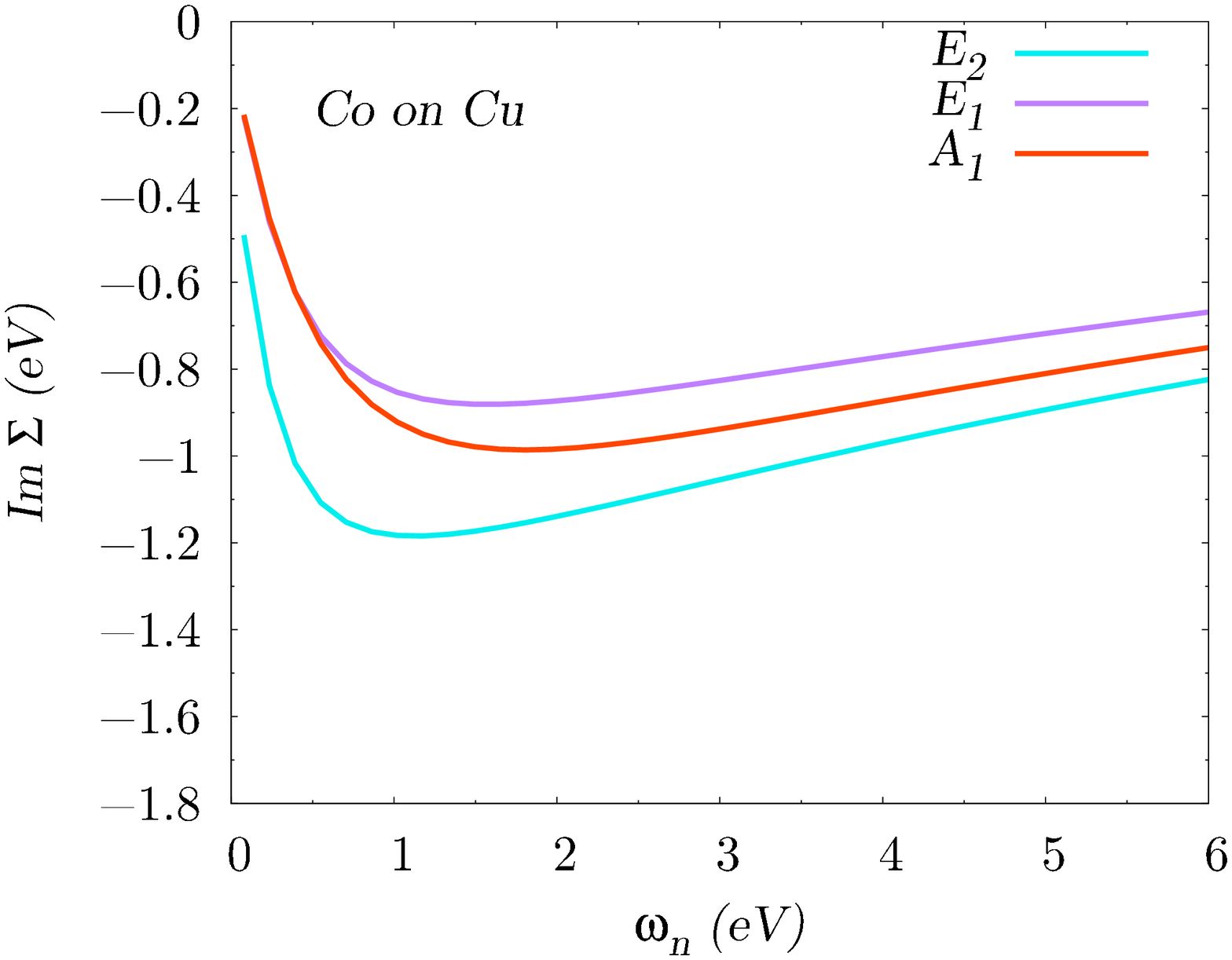}
\end{minipage}
\caption{Orbitally resolved self-energies for Co in Cu (upper panel) and Co on Cu (lower panel). From left to right , $\Im\,\Sigma(i \omega)$ is shown for the chemical potentials $\mu=26,27,28$\,eV (Co in Cu) and $\mu=27,28,29$\,eV (Co on Cu) .}%
\label{fig:CoCu_Sw_mu_b40}%
\end{figure*}

We now compare these relations to the frequency and the temperature dependence of $\Im \Sigma(T,i\omega_n)$ obtained from our QMC calculations. 
At the lowest accessible temperature, $T=0.025$\,eV, we obtained the Matsubara self-energies depicted for Co in and on Cu in Fig. \ref{fig:CoCu_Sw_mu_b40}. 

In both systems $|\text{Im}\Sigma|$ clearly decreases as $\omega_n\to 0$, for all orbitals except for the $E_2$ orbitals of Co on Cu at $\mu=27$\,eV. This is clearly different from the diverging $\Sigma(i\omega_n)\sim\frac{1}{i\omega_n}$, expected for the localized moment of an isolated atom.
For Co in bulk Cu, the $e_g$ and $t_{2g}$ orbitals exhibit very similar self-energies, whose low-energy behavior is consistent with the form expected for a Fermi-liquid (Eq.~\ref{eq:Im_Sigma_iomega}). For Co on Cu, the self-energies differ considerably between the different orbitals with the $E_1$ orbitals being least correlated and the $E_2$ orbitals exhibiting the largest self-energies at low frequencies. Our results indicate that a Fermi liquid develops in all Co orbitals, also here, although the Kondo temperature appears to be orbital dependent.

\subsection{Estimation of $T_K$ from QMC}
\label{sec:estimationoftk}
To define a  Kondo temperature scale even in cases without well defined local moment at intermediate temperatures, we define $T_K$ through the width of the quasi-particle resonance in the single particle spectral function near $E_F$, which is measured in STM experiments.

\begin{figure}
\includegraphics[width=9.0cm]{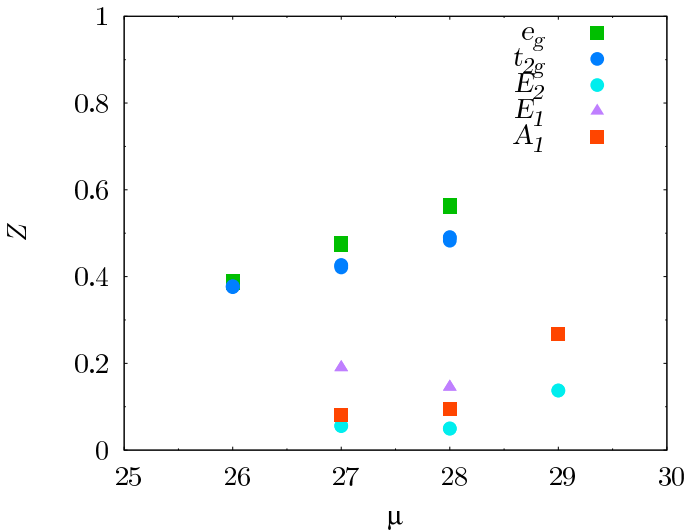}

\caption{(Color online)
Orbitally resolved quasi-particle weight of Co in Cu and Co on Cu at temperature $T=0.025$\,eV.
}
\label{fig:z}
\end{figure}

In our QMC simulations we  determine $T_K$ from the quasi-particle weight $Z$. The simulations yield the self-energy at the Matsubara frequencies $\omega_n=\frac{(2n+1)\pi}{\beta}$. Analytical continuation of Eq.~(\ref{eq:def_Z}) yields
\begin{equation}
\label{eq:z}
Z \approx \left(1-\left.\frac{\partial\text{Im}\Sigma(i\omega_n)}{\partial i\omega_n}\right|_{\omega_n=0} - \text{Re}\Delta'(0) \right)^{-1}
\end{equation}
and we use Eq.~(\ref{eq:Im_Sigma_iomega}) to evaluate the derivative. In Fig.~\ref{fig:z} we show the quasi-particle weight of Co in Cu and Co on Cu for the different types of orbitals as a function of the chemical potential $\mu$. 
The values of degenerate orbitals agree within an accuracy of $10^{-2}$ to which precision we are listing them in Tab. \ref{tab:tk} for $T=0.025$\,eV. The systems whose spin is closest to $S=1$ are found to have the lowest values of Z. Co in Cu clearly has higher quasi-particle weights compared to Co on Cu. As we will see, this results in a higher Kondo temperature $T_K$, which is also confirmed experimentally. In experiments using STM measurements the Kondo temperature has been found to be $T_K=655K \pm 155 K =0.056 \pm 0.013$\,eV for Co in Cu \cite{Wenderoth11} and $T_K\approx 54\pm 5\,{\rm K}=0.0046 \pm 0.0005$\,eV for Co on Cu.\cite{Manoharan_00,Kern02,Neel_08,Wenderoth11}

Following Hewson's derivation of a renormalized perturbation theory of the Anderson model \cite{Hewson:2005p1244} we use as a definition for the Kondo temperature:
\begin{equation}
\label{eq:tk}
T_K=-\frac{\pi}{4}Z\Im\Delta(0).
\end{equation}

The values computed according to Eq.~(\ref{eq:tk}) are listed in Tab. \ref{tab:tk} for all impurity energy levels $E_{\alpha}+\mu$ at temperature $T=0.025$\,eV. As for the quasi-particle weights we find the lowest values of $T_K$ for $\mu=26$\,eV (Co in Cu) and $\mu=28$\,eV (Co on Cu). Averaging over orbitals we obtain $T_K=0.118$\,eV (Co in Cu, $\mu=26$\,eV, $T=0.025$\,eV) and $T_K=0.016$\,eV (Co on Cu, $\mu=28$\,eV, $T=0.025$\,eV) with a ratio of $T_K^{\text{IN}}/T_K^{\text{ON}}=7.4$. This large difference between the two Kondo temperatures is in fair agreement with experiments, where a ratio of $T_K^{\text{IN}}/T_K^{\text{ON}}=12.1$ has been found.\cite{Wenderoth11} 
As discussed in Section \ref{scalepic} the physical quantities which determine the Kondo temperature scale enter in the argument of an exponential function, suggesting that a comparison of the logarithms $\log(T_{\rm exp})/\log(T_K)$ is more appropriate when judging the predictive power of our first-principles simulations. We find $\log(T_{\rm exp})/\log(T_K)=1.4$ (Co in Cu, $\mu=26$\,eV, $T=0.025$\,eV) and  $\log(T_{\rm exp})/\log(T_K)=1.3$ (Co on Cu, $\mu=28$\,eV, $T=0.025$\,eV).

\begin{table}[b]
\caption{
Kondo temperatures $T_K$ computed from the quasi-particle weight $Z$ and the hybridization function $\Delta(\omega)$ according to Eq. (\ref{eq:tk}) at the lowest simulation temperature $T=0.025$\,eV. The experimental Kondo temperatures are $T_K=0.056\pm0.013$\,eV (Co in Cu) and $T_K=0.0046\pm 0.0002$\,eV (Co on Cu).\cite{Wenderoth11,Kern02} \label{tab:tk} }
{\footnotesize
\begin{tabular*}{\columnwidth}{@{\extracolsep{\fill}} l c c c c c c}
\hline
\hline
System & $\mu$ (eV) & $E_i$ &   -Im($\Delta(0)$) (eV)  &  $Z$ & $T_K$ (eV)\\
\hline
Co in Cu &$26$ & $t_{2g}$ &  $0.43 \pm 0.01$ &0.38 & 0.13   \\
Co in Cu &$26$ & $e_g$ & $0.340 \pm 0.009$ &0.39 & 0.10   \\
\hline
Co in Cu &$27$ & $t_{2g}$ & $0.43 \pm 0.01$ &0.42 & 0.14   \\
Co in Cu &$27$ & $e_g$ & $0.340 \pm 0.009$ &0.47 & 0.12   \\
\hline
Co in Cu &$28$ & $t_{2g}$ & $0.43 \pm 0.01$ &0.48 & 0.16   \\
Co in Cu &$28$ & $e_g$ & $0.340 \pm 0.009$ &0.56 & 0.15   \\
\hline
\hline
Co on Cu & $27$& $E_2$ & $0.124 \pm 0.002$ &0.06 & 0.006   \\
Co on Cu & $27$& $E_1$ & $0.226 \pm 0.002$ &0.19 & 0.03   \\
Co on Cu & $27$& $A_1$ & $0.197 \pm 0.001$ &0.08 & 0.01   \\
\hline
Co on Cu & $28$& $E_2$ & $0.124 \pm 0.002$ &0.05 & 0.005   \\
Co on Cu & $28$& $E_1$ & $0.226 \pm 0.002$ &0.15 & 0.03   \\
Co on Cu & $28$& $A_1$ & $0.197 \pm 0.001$ &0.10 & 0.01   \\
\hline
Co on Cu & $29$& $E_2$ & $0.124 \pm 0.002$ &0.14 & 0.01   \\
Co on Cu & $29$& $E_1$ & $0.226 \pm 0.002$ &0.26 & 0.05   \\
Co on Cu & $29$& $A_1$ & $0.197 \pm 0.001$ &0.27 & 0.04   \\
\hline
\hline
\end{tabular*}
}
\end{table}

For both systems our computed Kondo temperatures are higher than the experimentally determined values. This can be either due to the neglect of spin-orbit coupling effects which lifts degeneracies and narrows the low energy resonances or due to the Coulomb interactions being larger than  the $U=4$\,eV assumed here.

\begin{figure*}
\includegraphics[width=8cm]{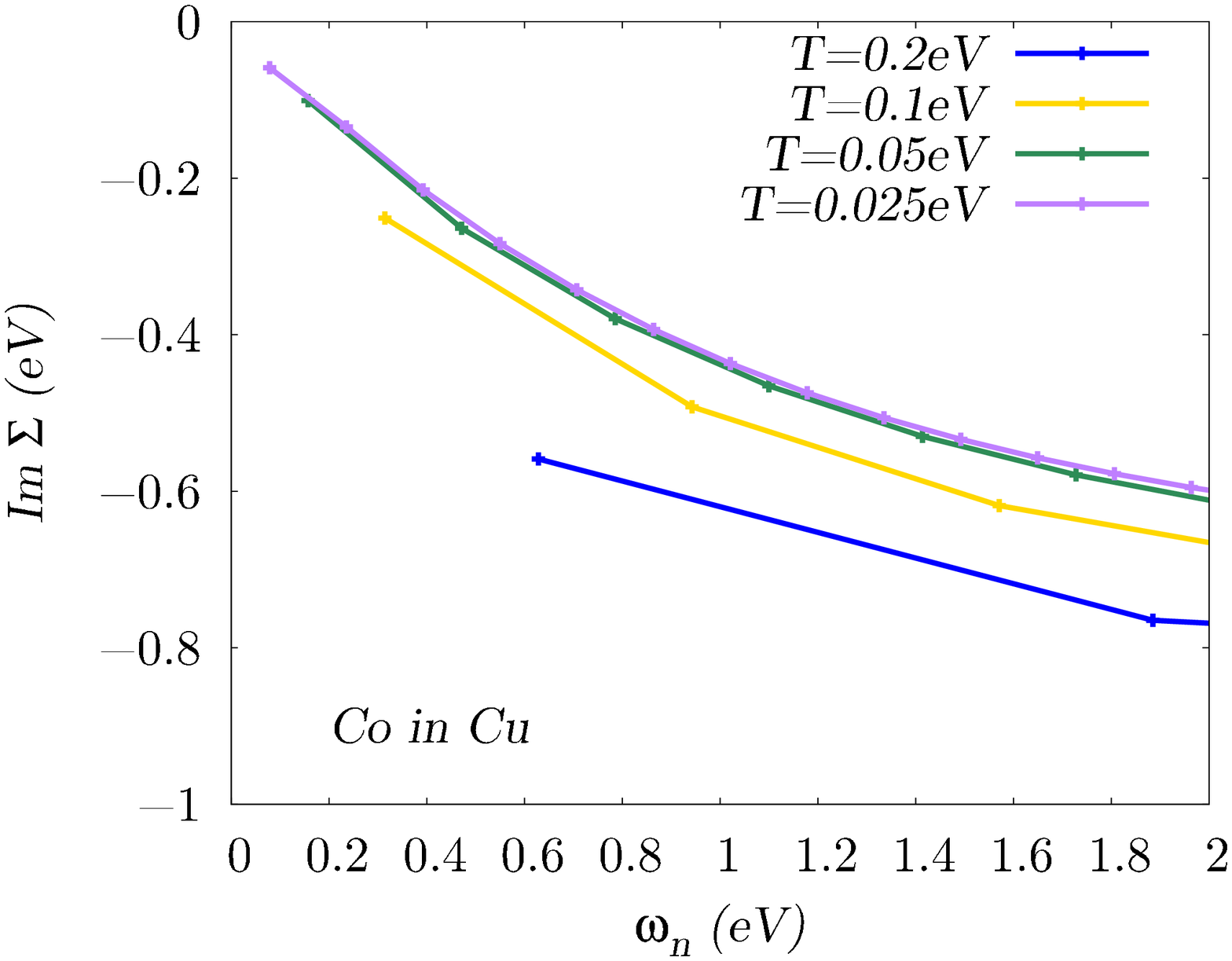}
\includegraphics[width=8cm]{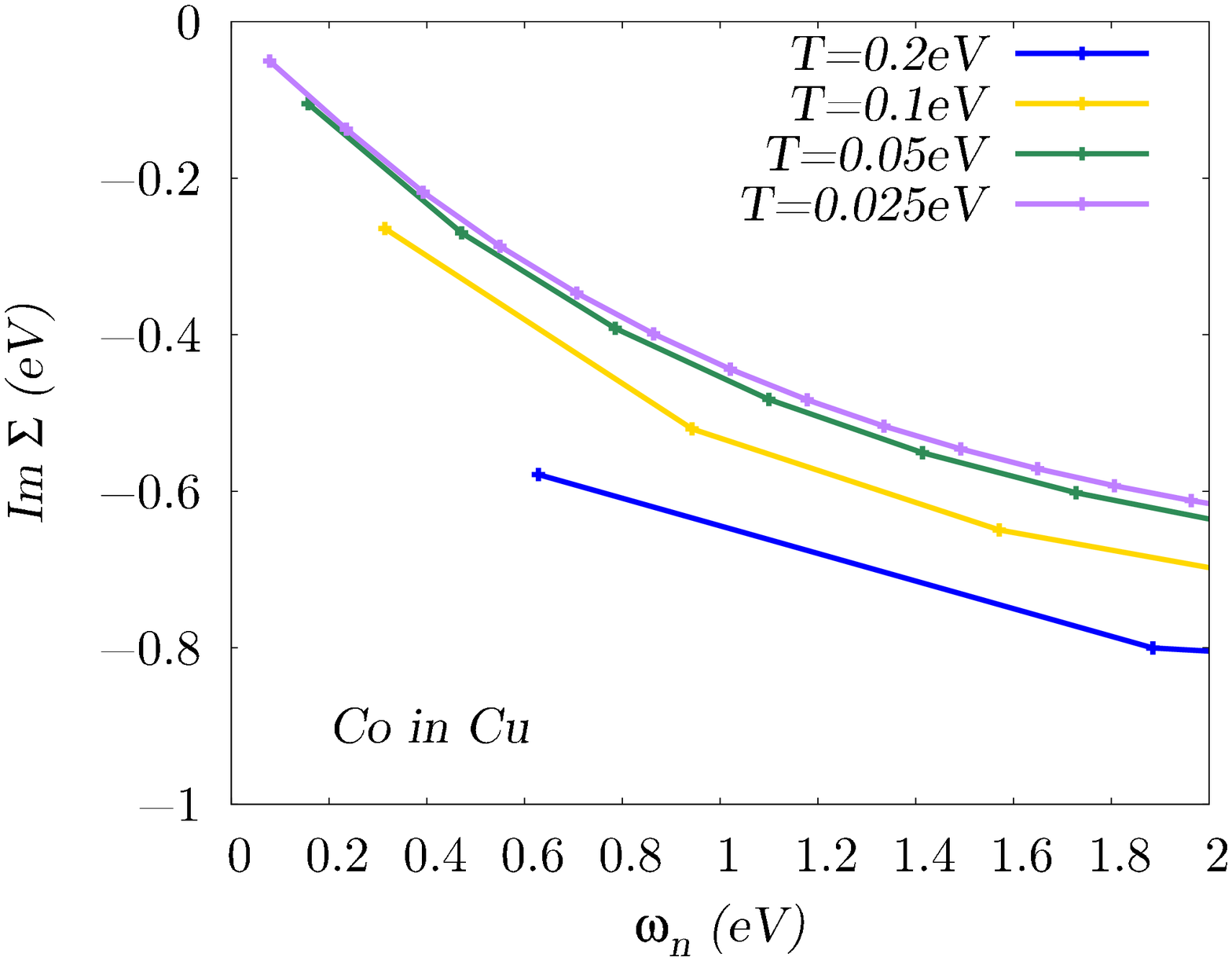}

\includegraphics[width=5.5cm]{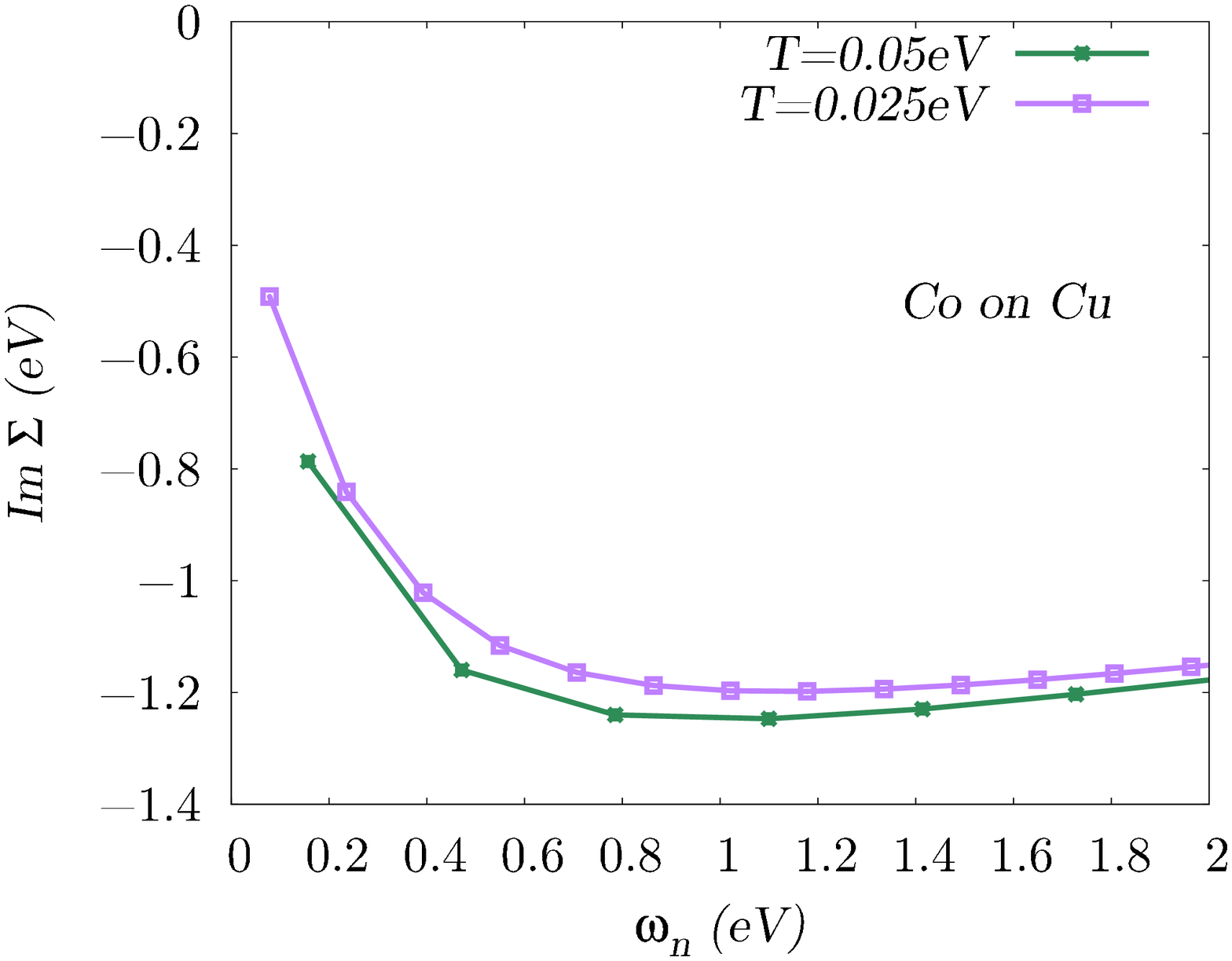}
\includegraphics[width=5.5cm]{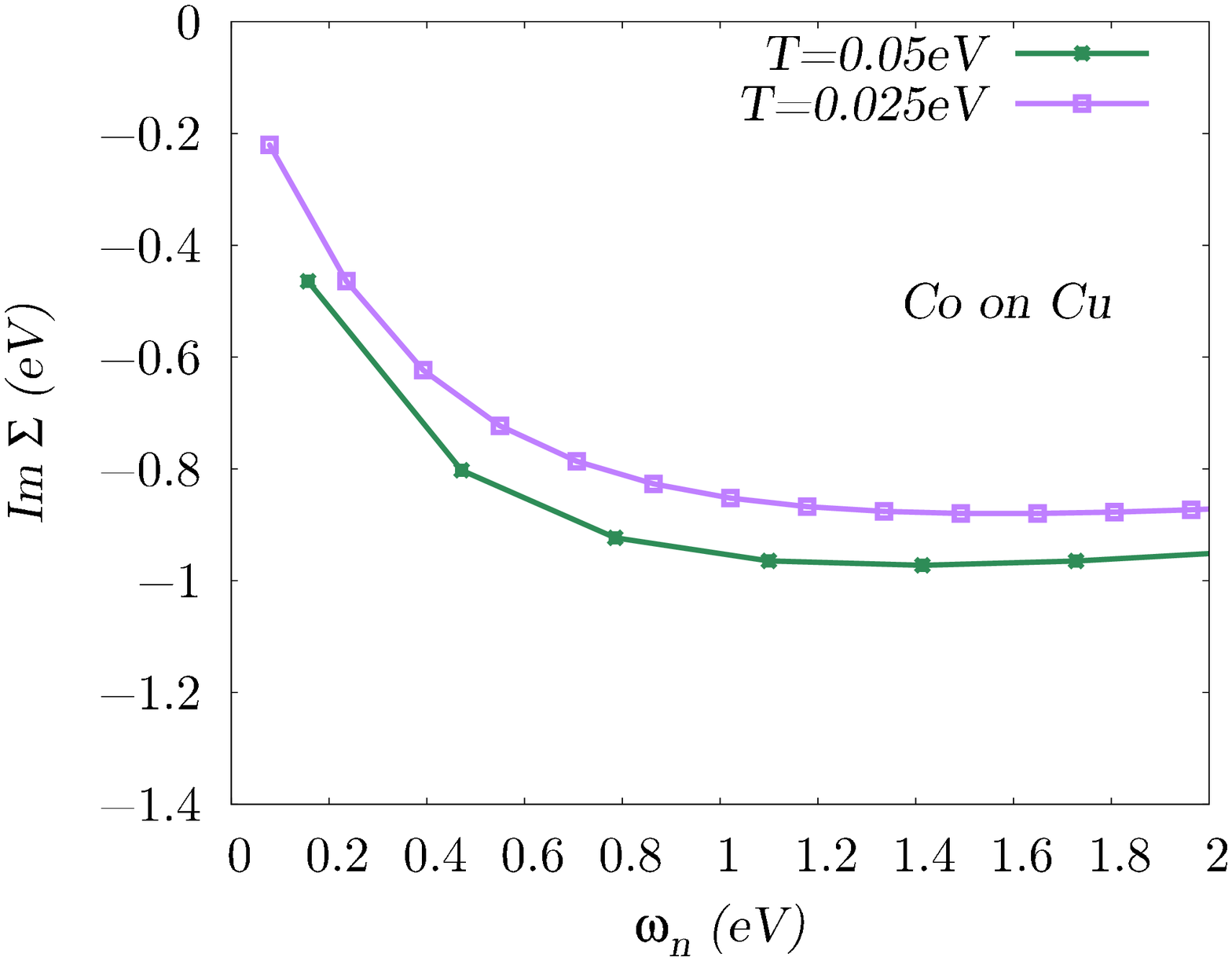}
\includegraphics[width=5.5cm]{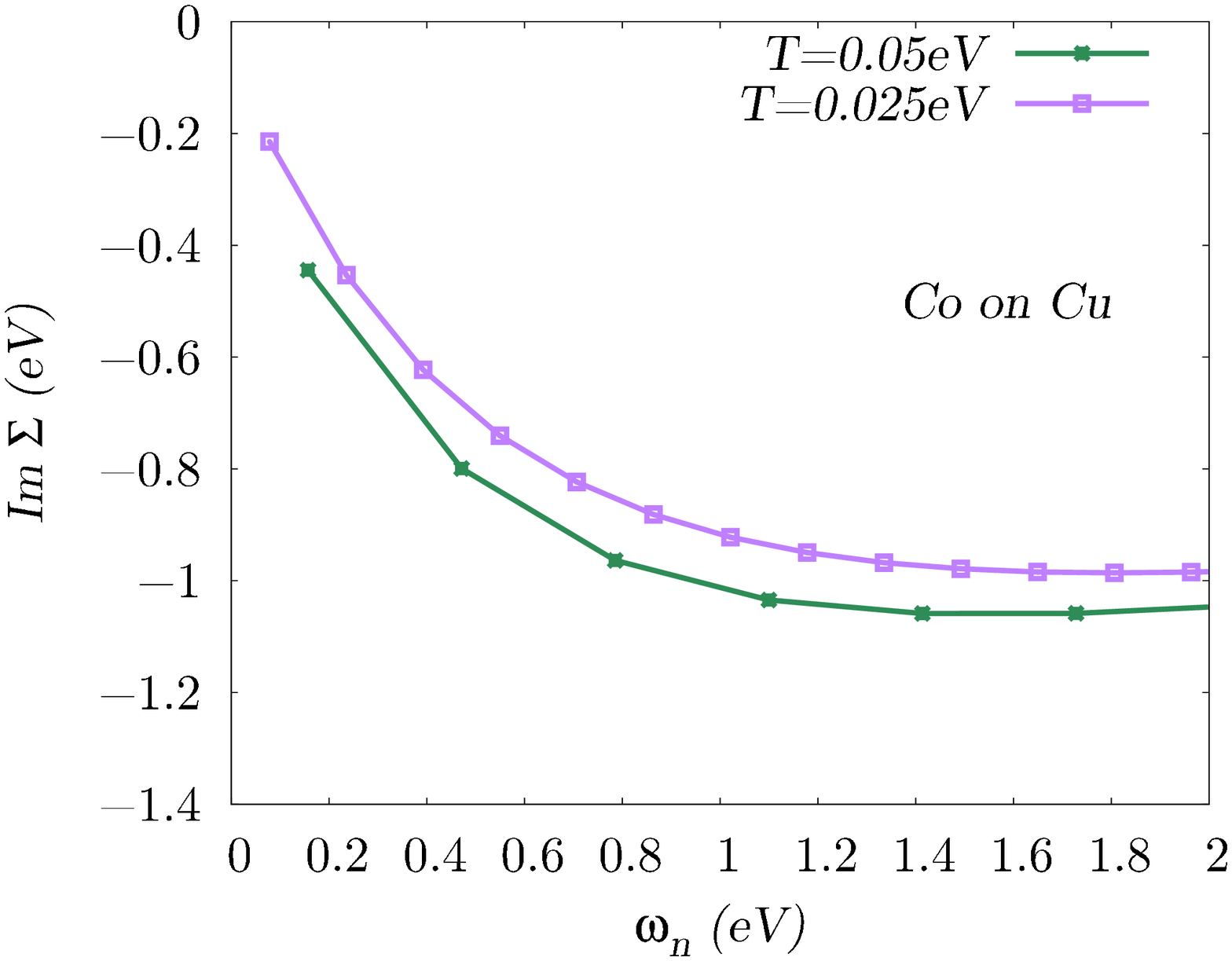}
\caption{(Color online) Upper panel: Im$(\Sigma)$ of Co in Cu for $\mu=28$\,eV at orbital energies $E_i=-0.288$\,eV ($d_{xy},d_{yz},d_{xz}$) and $E_i=-0.44$\,eV ($d_{z^2},d_{x^2-y^2}$) (from left to right). Lower panel:
Im$(\Sigma)$ of  Co on Cu for $\mu=29$\,eV at orbital energies $E_i=-0.12$\,eV  ($d_{xy},d_{x^2-y^2}$), $E_i=-0.39$\,eV  ($d_{yz},d_{xz}$) and $E_i=-0.34$\,eV ($d_{z^2}$) (from left to right).}
\label{fig:Sw_beta_coincu_cooncu}
\end{figure*}

The temperature dependence of the self-energy allows for an alternative test whether and at which energy scale a Fermi liquid emerges. This is illustrated for Co in and on Cu in Fig. \ref{fig:Sw_beta_coincu_cooncu}.
We find an almost temperature independent behavior of $\Im\,\Sigma(T,i\omega_n)$ for Co in Cu if $T<0.05\,$eV, which provides an estimate of $T_K\approx 0.05\,$eV. In the case of Co on Cu $\Im\,\Sigma(T,i\omega_n)$ still evolves as one lowers the temperature from $T=0.05\,$eV to $T=0.025$\,eV, which suggests a lower Kondo temperature.

\begin{figure}%
\begin{minipage}{.49\linewidth}
\includegraphics[width=.99\columnwidth]{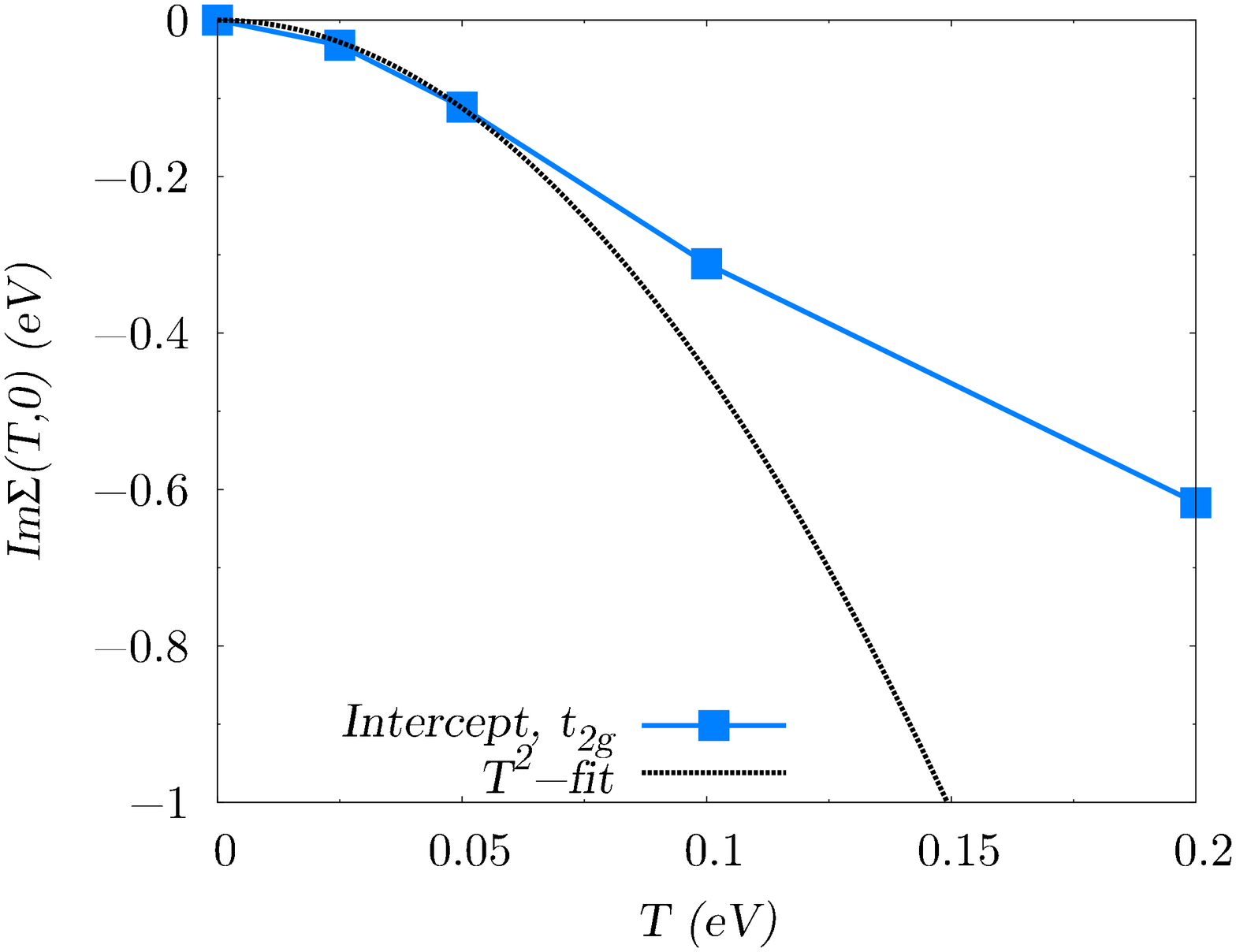}
\end{minipage}
\begin{minipage}{.49\linewidth}
\includegraphics[width=.99\columnwidth]{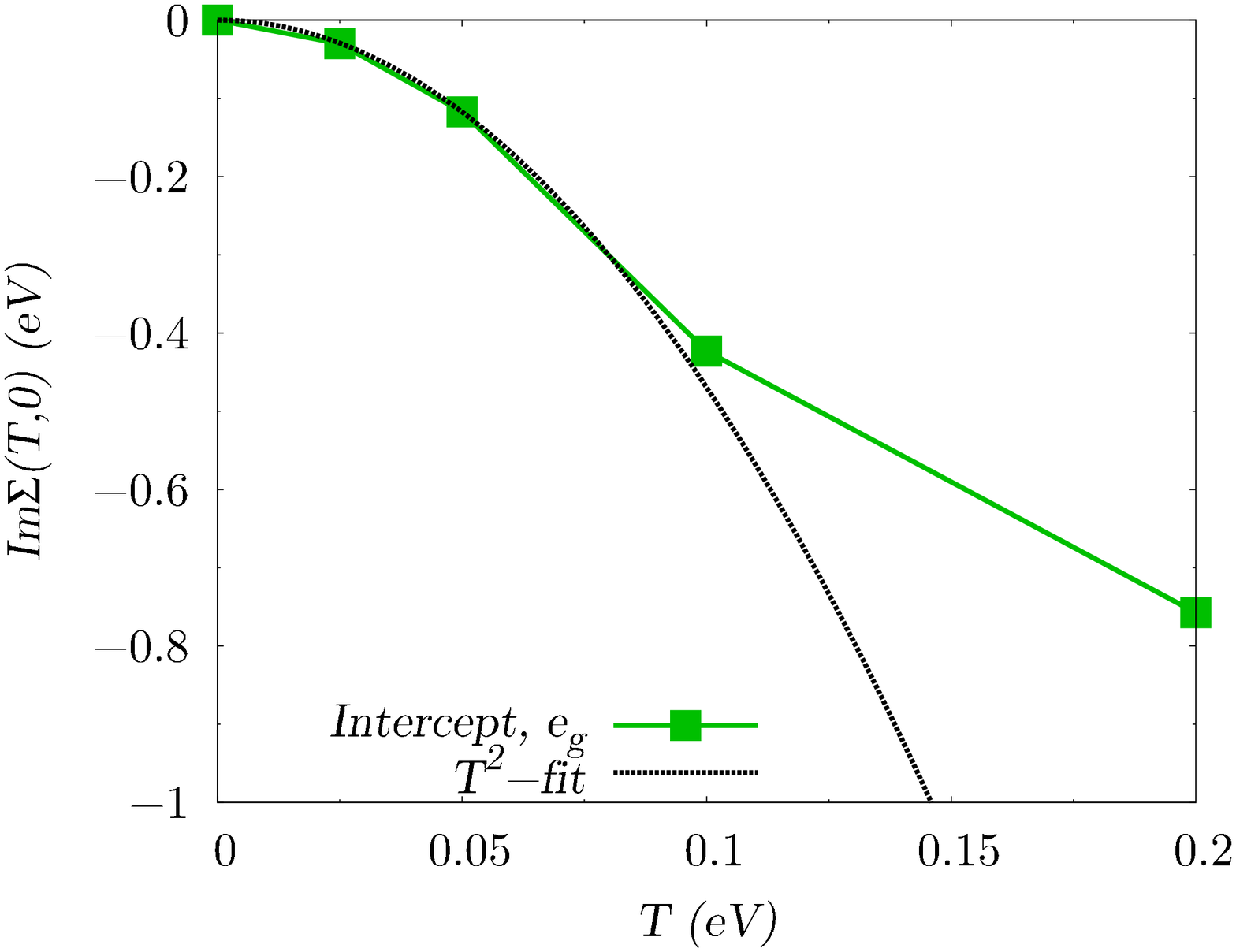}
\end{minipage}
\caption{(Color online) Temperature dependent quasi-particle lifetimes for Co in Cu at $\mu=26$\,eV for the $t_{2g}$ (left) and $e_g$ orbitals (right).
}%
\label{fig:intercept}%
\end{figure}

According to Eq.~(\ref{eq:Im_Sigma_iomega}), the linear extrapolation of $\Im \Sigma(T,i \omega)$ to $i\omega_n\rightarrow 0$ yields the inverse lifetime $\hbar\tau^{-1}=\Im \Sigma(T,0)$ of quasi-particles at the Fermi level if a Fermi liquid is formed. In Fig.~\ref{fig:intercept} we plot this extrapolation for Co in Cu, $\mu=26$\,eV and show $\Im \Sigma(T,0)$ as a function of temperature. We find a $\Im \Sigma(T,0)\sim T^2$-behavior for both sets of orbitals, which corroborates the formation of a Fermi-liquid in the $e_g$ and $t_{2g}$ orbitals at the lowest accessible temperatures.




\section{Discussion}
\label{scalepic}
Our QMC calculations showed that, for both  Co in and on Cu Fermi liquids involving all orbitals of the Co impurity form at low temperatures. We now want to understand these results on the basis of scaling arguments starting with (higher energy) charge fluctuations going to (lower energy) spin and orbital fluctuations.

\subsection{Charge fluctuations}
\label{sec:chargeflucuations}
With our values of the Coulomb interaction strength in the local Hamiltonian ($U=4$\,eV; $J=0.9$\,eV) the energies of removing ($E_-$) or adding ($E_+$) an electron to the impurity are $E_-\approx E_+\approx 2$\,eV. A first order expansion in the hybridization gives a qualitative estimate of the role of charge fluctuations:\cite{NozieresBlandin_1980} The norm of the admixtures of $d^7$ and $d^9$ configurations to a predominantly $d^8$ ground state of the impurity is approximately $\mathcal{N}_{n\neq 8}=-\frac{1}{\pi}\Im \Delta(0)\frac{10}{U/2}$. For Co in Cu, $-\Im\,\Delta(0)\approx 0.4$\,eV leads to $\mathcal{N}_{n\neq 8}\approx 0.6$. The hybridization of Co on Cu is about twice smaller, $-\Im\,\Delta(0)\approx 0.2$\,eV, yielding a correspondingly smaller weight of non-$d^8$ configurations $\mathcal{N}_{n\neq 8}\approx 0.3$. 

\begin{figure}
\begin{minipage}{1.\linewidth}
\includegraphics[width=\linewidth]{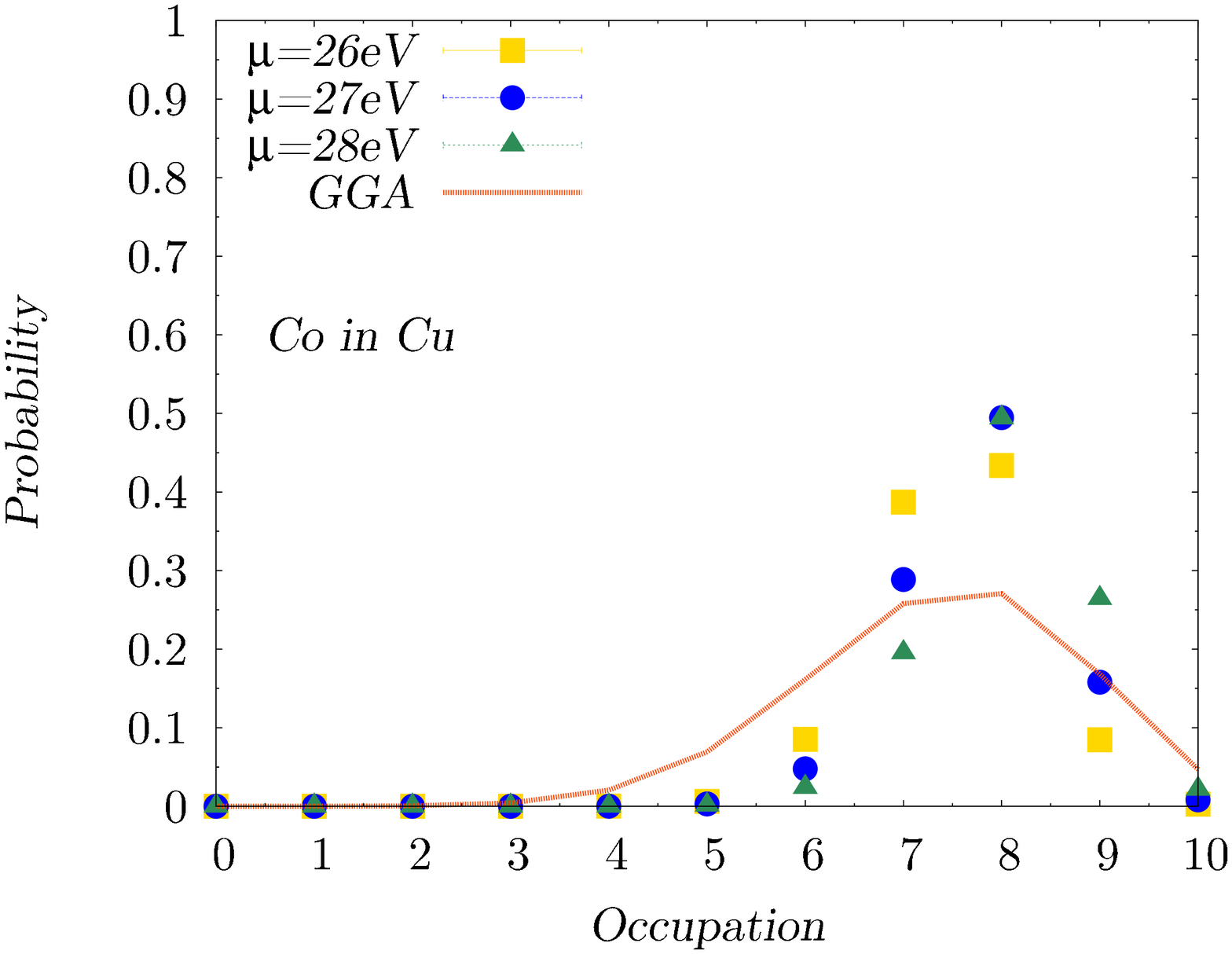}
\end{minipage}
\begin{minipage}{1.\linewidth}
\includegraphics[width=\linewidth]{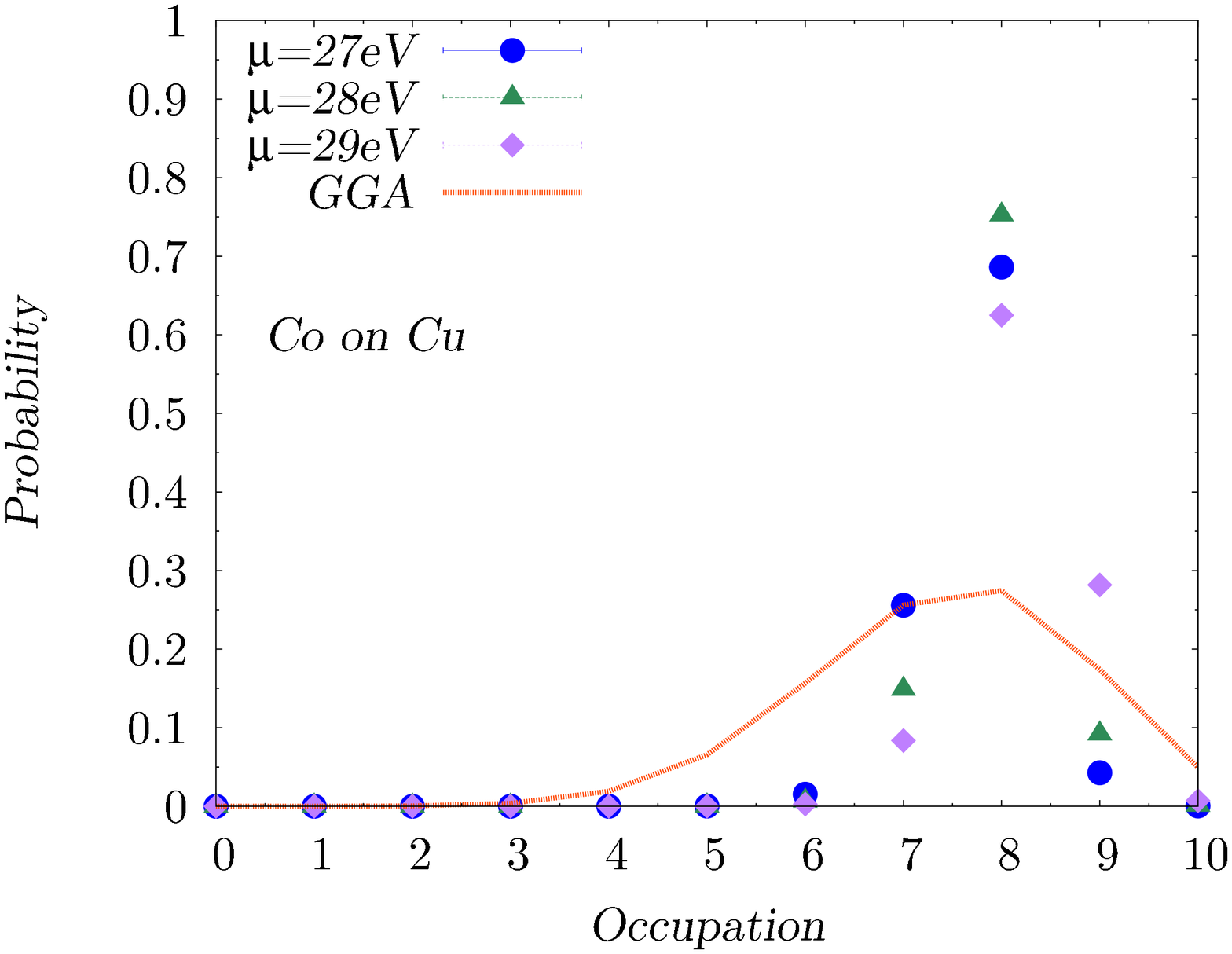}
\end{minipage}
\caption{(Color online)
Top: Occupation statistics of Co in Cu for $\beta=40$. Bottom: Occupation statistics of Co on Cu for $\beta=40$.
}
\label{fig:pn}
\end{figure}

Our QMC calculations allow to quantitatively measure the charge fluctuations. In Fig.~\ref{fig:pn} we plot the ``valence histogram" \cite{Haule:2007p693} for Co in/on Cu for different choices of the chemical potential and compare it to the ``valence histogram" of the Slater determinant built from the lowest GGA eigenstates. The histogram shows the weights which the eigenstates in the different charge sectors $n=0$, $1$, \ldots $10$ contribute to the partition function (via the outer trace in Eq.~(\ref{trace})). In all cases the local Coulomb interaction in the QMC simulation leads to a narrower distribution of the occupancies as compared to the GGA valence histograms. This effect is most pronounced in the case of Co on Cu (111), where the $d^8$ configuration clearly dominates over the $d^7$ and $d^9$ configurations. For Co in Cu, there are still noticeable correlations and the narrowing of the valance histogram as compared to the GGA case but the $d^8$ configuration contributes only about 50\% in the QMC simulations, with significant weight coming from the $d^7$, $d^9$ and even the $d^6$ and $d^{10}$ configurations. The measured values for $\mathcal{N}_{n\neq 8}$ ($\approx 0.5$ for Co in Cu and $\approx 0.3$ for Co on Cu) agree surprisingly well with the simple perturbative estimate.

A Schrieffer-Wolff decoupling of ionized impurity states to discuss the low energy physics can be justified if the weight of  non-$d^8$ configurations is $\mathcal{N}_{n\neq 8}\ll 1$. Hence, an intermediate energy scale of well formed (unscreened) fluctuating spin or orbital moments at frozen impurity valency might be defined in the case of Co on Cu (111) but clearly not in the case of Co in Cu.

\subsection{Spin and orbital fluctuations}
\label{sec:spinandorbitalflucations}

While for Co in Cu charge, spin and orbital fluctuations will be present down to lowest energies, for Co on Cu only fluctuations of the orbital and the spin degree of freedom are  expected to dominate the low energy physics. To further investigate to which extent orbital and spin fluctuations might determine the low energy behavior of the impurity we estimate Kondo temperatures within simplified models and compare these estimates to our QMC calculations as well as experiments.

Assuming well-defined magnetic moments ($\mathcal{N}_{n\neq 8}\ll 1$) a scaling analysis (cf. Ref. \onlinecite{NozieresBlandin_1980}) allows to estimate
the Kondo scale analytically in simplified situations: If neither Hund's rule coupling nor crystal field splitting or any other symmetry breaking terms were present the spin and the orbital degrees of freedom of the Co impurities could fluctuate freely and independently. This would lead to a Kondo temperature \cite{NozieresBlandin_1980} $T_K\sim D_0\exp\left[-1/2\mathcal{N}_{n\neq 8}\right]$, where $D_0\sim\min(E_\pm,\Lambda)$ is related to the impurity charging energies ($E_\pm$) and the electronic bandwidth $\Lambda$. $D_0$ is on the order of several eV. With $D_0=U/2=2\,$eV we estimate $T_K\approx 0.4 D_0=0.9\,$eV in the case of Co in Cu and $T_K\approx 0.2D_0=0.4\,$eV in the case of Co on Cu. This is in both cases (at least) an order of magnitude larger than the Kondo temperatures obtained from QMC and the experimentally measured Kondo temperatures.

The opposite limit is given by the case with strong Hund's rule coupling and supressed orbital fluctuations. Without orbital fluctuations, but still disregarding the Hund's coupling $J$, the Kondo temperature reads \cite{NozieresBlandin_1980,HewsonBook} $T_K\sim D_0\exp\left[\frac{\pi U}{8\Im\,\Delta(0)}\right]=D_0\exp\left[-\frac{5}{2\mathcal{N}_{n\neq 8}}\right]$, which would lead to $T_K=0.02D_0\approx 0.04$\,eV for Co in Cu. The Hund's rule coupling reduces the Kondo temperature \cite{Coleman_PRL09} further to $T_K^*=T_K(T_K/J_H S)^{2S-1}$. With $S=1$ and $J_H=0.9$\,eV this would lead to $T_K^*\approx 0.002\,$eV. For Co on Cu, the assumption of an orbital singlet yields $T_K=0.0004 D_0\approx 0.001$\,eV and Hund's rule coupling further reduces the Kondo temperature to $T_K^*\approx 1\,\mu$eV. This limit thus yields Kondo temperatures which are orders of magnitude smaller than those obtained in our QMC calculations as well as the Kondo temperatures measured experimentally for Co in and on Cu.

It is thus the successive locking of the impurity electrons to a large spin by the Hund's rule coupling and the \textit{partial} freezing out of orbital fluctuations that determines the on-set of Fermi liquid behavior and the Kondo temperature in realistic systems like Co in or on Cu.

With this in mind it is instructive to analyze the influence of a static crystal field on the energy spectrum of otherwise isolated Co atoms.
Without crystal fields, in a $d^8$ configuration our local Coulomb interaction ($U=4$\,eV; $J=0.9$\,eV) yields a 21 fold degenerate $L=3, S=1$ ground state which is separated from the $L=2,S=0$ multiplet by an energy of $E_{L=2,S=0}=1.3$\,eV. 
This is clearly larger than the crystal field acting on the Co impurities (Fig. \ref{fig:DFT_hybridization}): The cubic crystal field (evaluated at the Fermi level) of Co in Cu leads to the $e_g$ states being $0.18$\,eV higher in energy than the $t_{2g}$ states. In this crystal field, the resulting $d^8$ ground state is an orbital singlet. Excitations to higher crystal field split states require energies on the order of $0.2$\,eV. This is larger but comparable in order of magnitude to the Kondo temperatures obtained in experiments and simulation. However, fluctuations to these higher crystal field split states must be taken into account to explain the characteristic temperature of the low energy Fermi liquid formed at Co impurities in Cu.

In our model of Co on Cu, the static crystal fields also lift the degeneracy of the ground state multiplet but a double degeneracy in the orbital space remains. Excitations to higher crystal field split states require $0.03-0.08$\,eV. 
In this model, even the ground state multiplet allows for fluctuations of the orbital degree of freedom.


\begin{figure}
\begin{minipage}{0.49\linewidth}
\includegraphics[width=\linewidth]{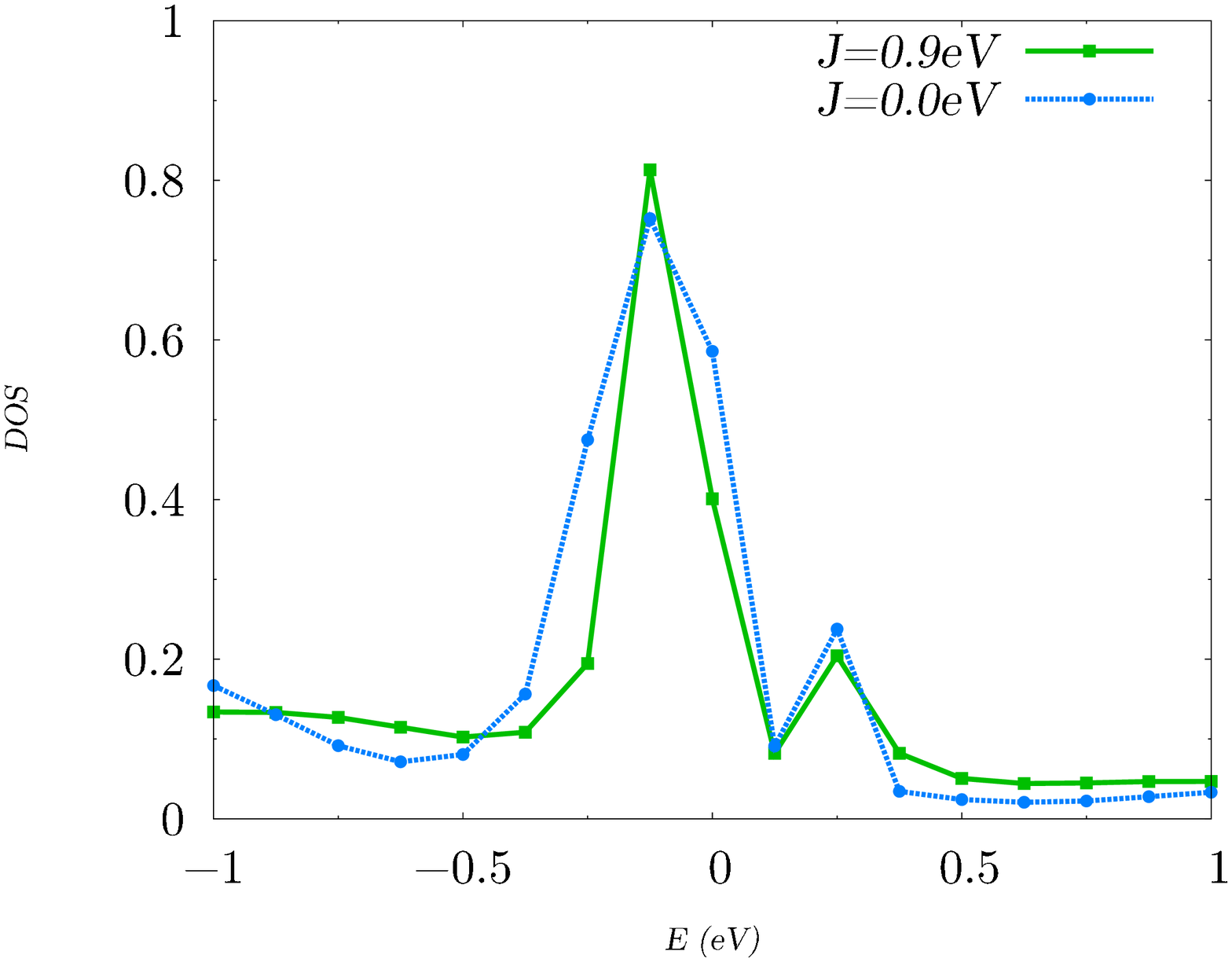}
\end{minipage}
\begin{minipage}{0.49\linewidth}
\includegraphics[width=\linewidth]{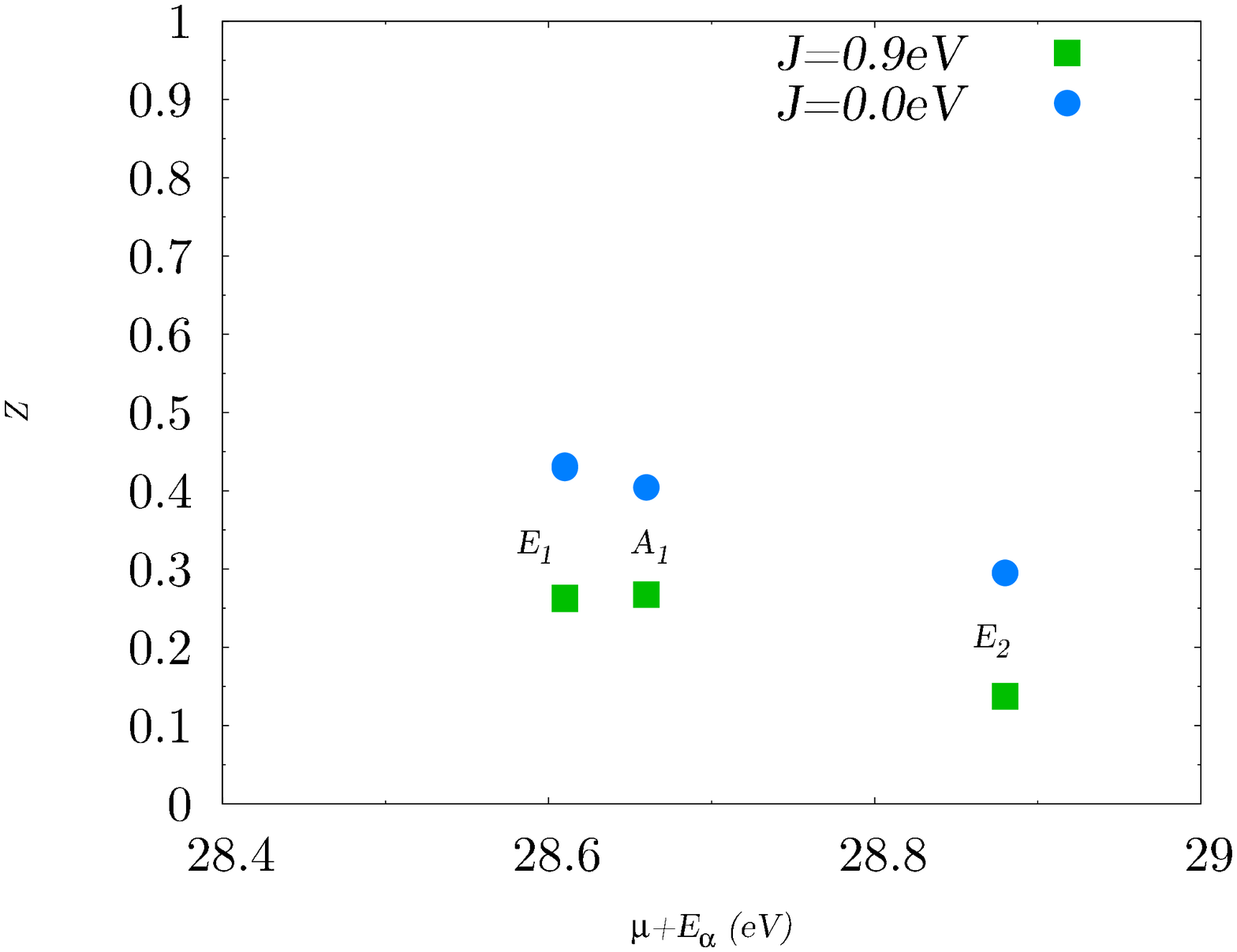}
\end{minipage}
\caption{(Color online) Comparison of the DOS (left) and quasi-particle weight Z (right) of Co on Cu at $U=4$\,eV for the two values $J=0.0$\,eV and $J=0.9$\,eV.}%
\label{fig:j0}%
\end{figure}

In order to examine the effect of constraining orbital fluctuations we consider the case of Co on Cu which exhibits a strong reduction of the quasi-particle peak compared to the GGA spectral function representing the $U=0$, $J=0$ case. Turning off the Hund's coupling $J$ allows the orbital and spin degrees of freedom to fluctuate more freely and given the scaling considerations should result in a higher Kondo temperature as well as a broader quasi-particle peak. We study the effect of $J=0$ for Co on Cu, $T=0.025$\,eV, $\mu=29$\,eV and present the comparison of the quasi-particle weight and peak in Fig.~\ref{fig:j0}. In line with our statement that the Kondo temperature is determined by the locking of the impurity electrons to a larger spin and possible restrictions of the orbital fluctuations, we find a broadening of the quasi-particle peak and increase of the quasi-particle weight $Z$ as $J\to 0$.

\section{Conclusions}
\label{sec:conclusions}
For Co in and on Cu we found that a Fermi liquid is formed at low $T$ involving all impurity $d$-orbitals. The example of Co on Cu shows that the characteristic temperature, $T_K$, associated with the onset of Fermi liquid behavior can differ between the impurity orbitals. The comparison of our QMC calculations and scaling arguments further demonstrates that fluctuations in the orbital degree of freedom and Hund's rule coupling are crucial in determining $T_K$ in realistic systems. This is well beyond the physics of simple ``spin-only'' models. The understanding of magnetic nanostructures based on 3$d$ adatoms on surfaces as well as magnetic impurities in bulk metals require us to account for the orbital degrees of freedom. Dynamical mean field theory provides a link between quantum impurity problems and extended lattices of atoms which are subject to strong electron correlations. 

It remains thus a future challenge to understand how the orbital degree of freedom controls the quenching of magnetic moments and eventually the formation of low energy Fermi liquids in realistic extended correlated electron systems as well as magnetic nanostructures. 

\acknowledgements{We are grateful to S. Brener, G. Czycholl, M. Katsnelson, A. Rosch and M. Sigrist for useful discussions and thank L. Boehnke for sharing his python MaxEnt code with us. This work was supported by SFB 668 (Germany), SNF grant PP002-118866, the Cluster of Excellence `Nanospintronics'' (LExI Hamburg) is acknowledged. The calculations have been performed at HLRN (Germany) and the Swiss National Supercomputing Centre using the ALPS libraries.\cite{alps}}

\bibliography{CoCu_Kondo}

\end{document}